\documentclass{style}
\usepackage{graphicx}
\usepackage{color} 

\newcommand{\be}{\begin{equation}}
\newcommand{\ee}{\end{equation}}
\newcommand{\bea}{\begin{eqnarray}}
\newcommand{\eea}{\end{eqnarray}}
\newcommand{\br}{{\bf r}}

\begin{document}

\title{Isovector spin-singlet (T=1, S=0) and isoscalar spin-triplet (T=0, S=1) 
pairing interactions and spin-isospin response}
\author{
  H. Sagawa\email{sagawa@ribf.riken.jp}\\
  \it RIKEN, Nishina Center, Wako, 351-0198, Japan\\
  and \\
\it  Center for Mathematics and Physics, University of Aizu, \\
\it  Aizu-Wakamatsu,
Fukushima 965-8560, Japan\\
  C. L. Bai\email{bclphy@scu.edu.cn} \\
  \it School of Physical Science and Technology, Sichuan University, Chengdu 610065, China\\
  G. Col\`o\email{colo@mi.infn.it}\\
  \it Dipartimento di Fisica, Universit$\grave{a}$ degli Studi di
Milano \\
 and \\
\it  INFN, Sezione di Milano, 20133 Milano, Italy
}


\date{}

\maketitle

\begin{abstract}
We review several experimental and theoretical advances that emphasise common aspects 
of the study of spin-singlet, $T=1$, and spin-triplet, $T=0$, pairing correlations in nuclei. 
We first discuss several empirical evidences of the special role played by the $T=1$ 
pairing interaction. In particular, we show the peculiar features of the nuclear pairing 
interaction in the low density regime, and possible outcomes such as the BCS-BEC crossover 
in nuclear matter and, in an analogous way, in loosely bound nuclei. We then move to the
competition between $T=1$ and $T=0$ pairing correlations. 
The effect of such competition on the low-lying spectra is studied
in $N=Z$ odd-odd nuclei by using a three-body model; in this case, it is shown that the inversion 
of the $J^{\pi}=0^+$ and $J^{\pi}=1^+$ states near the ground state, and the strong magnetic 
dipole transitions between them, can be considered as a clear manifestation of strong 
$T=0$ pairing correlations in these nuclei. The effect of $T=0$ pairing correlations is
also quite evident if one studies charge-changing transitions.
The Gamow-Teller (GT) states in $N=Z+2$ nuclei are studied here by using 
self-consistent Hartree-Fock Bogoliubov (HFB) plus Quasi-particle Random Phase Approximation 
(QRPA) calculations in which the $T=0$ pairing interaction is taken into account. Strong GT states 
are found, near the ground state of daughter nuclei; these are compared 
with available experimental data from charge-exchange reactions, and such
comparison can pinpoint the value of the strength of the $T=0$ interaction.
Pair transfer reactions are eventually discussed: while two-neutron transfer 
has been long proposed as a tool to measure the $T=1$ superfluidity in the nuclear ground states, 
the study of deuteron transfer is still in its infancy, despite its potential interest in 
revealing effects coming from both $T=1$ and $T=0$ interactions.
We also point out that the reaction mechanism may mask the strong pair transfer amplitudes 
predicted by the HFB calculations, because of the complexity arising from 
simultaneous and sequential pair transfer processes. 
\end{abstract}
  

\section{Introduction --- superfluidity in nuclei}
\label{Sect.01}

The effective nucleon-nucleon interactions 
that are employed within self-consistent mean-field approaches have recently reached
a high level of sophistication, and have become
quite successful in describing many nuclear properties. Though they can be based on different kinds 
of ansatz, 
central, spin-orbit, and tensor terms show up always. In open-shell nuclei the pairing interaction or,
in fact, its isovector ($T=1, S=0$) part, has been originally introduced to account for the 
odd-even binding
energy staggering, the gap in the excitation spectrum of even-even and odd-$A$ 
nuclei~\cite{Bohr1,Bohr2,Brink1}, the moment of inertia of deformed nuclei \cite{Bohr2} and also the fission barrier of actinide nuclei (cf. p. 158 of \cite{Brink1}).
In the literature, 
only the spin-singlet $T=1$ pairing has been mainly discussed in nuclear physics, 
since the large spin-orbit splitting prevents to couple 
a spin-triplet ($T=0$, $S=1$) pair in the ground state \cite{Bertsch2012,Sagawa2013}.  
Another reason for this is 
that the neutron excess along the stability line of 
the nuclear chart suppresses the proton-neutron pairing for medium-mass and heavy nuclei.
The recent availability of radioactive beams 
has opened up new opportunities
to measure structure properties 
of unstable nuclei along the $N=Z$ line, strongly enhancing 
the possibility to measure new properties of nuclei such as pairing 
correlations related with the spin-triplet $T=0$ pairing \cite{Tanimura12}. 
It is thus quite interesting and important to study the competition between the spin-singlet $T=1$ and the spin-triplet $T=0$ pairing 
interactions in $N\approx Z$ 
nuclei, and  
seek an experimental evidence for 
their competition in the energy spectra and the transition rates. 

One of the widely used mean-field approaches is based on zero-range Skyrme forces: 
Hartree-Fock (HF) plus Bardeen-Cooper-Schrieffer (BCS) equations~\cite{Vautherin,Ring} 
or Hartree-Fock-Bogoliubov (HFB) equations~\cite{Ring,Dob1}, which include
the pairing interaction, can be, and have been solved, to
study the ground state properties of the open-shell nuclei~\cite{Ben03,Stoi06,dug01,mar07}. 
On top of this ground-state solutions, the self-consistent Quasi-particle Random-Phase-Approximation (QRPA)  
has been adopted 
by many authors 
to study the collective 
excited states~\cite{Ring,Rowe,Terasaki,Engel,M01,Khan02,Paar03}.
  
The parameters that characterize the effective interactions like the Skyrme ones
can be fitted by using empirical
properties of uniform nuclear matter, as well as few ground-state (or sometimes excited state) 
properties of finite nuclei. 
However,  
some channels of the interactions are not well constrained, one of the clearest examples being 
the pairing interaction between protons and neutrons
in the isoscalar spin-triplet ($T=0$, $S=1$) channel. 
Indeed, there is no consensus on the observables
that can be directly related to such channel, and not yet unambiguous signatures of strong neutron-proton 
particle-particle correlations, despite several 
efforts \cite{Goodman1,Goodman2,Poves,Ga1,Ga2,Bertsch1,Bertsch2}.

It has been well known that one of the effects of the bare 
isoscalar spin-triplet force is to
give rise to the deuteron bound state. 
Although some speculations have been made about the relevance of a n-p pairing force in nuclei
with $N = Z$, there is not unambiguous evidence of its effects, 
let alone evidence of a p-n condensate. 
In Ref. \cite{Bertsch1}, it has been shown that ordinary bound
nuclei are dominated by spin-orbit effects but on the other hand, 
if such effects can be made less important and one considers either
very large nuclei or nuclei with low angular momentum orbitals
close to the Fermi surface, then n-p pairing does manifest itself 
strongly.

Low-lying states of nuclei having a neutron and a proton outside a relatively 
closed core may be good candidates to study n-p pairing, as far as the two particles lie
in the same orbital. However, collective states are probably better candidates to
extract more firm and general information. In this respect,
effects of isoscalar pairing may be present in
charge-exchange excitations and related phenomena. 
Indeed, in self-consistent Skyrme QRPA calculations it can be shown that the Gamow-Teller 
Resonance (GTR) is only sensitive to $T = 0$ pairing, while the Isobaric Analog Resonance 
(IAR) is only sensitive to $T = 1$ pairing \cite{Fr}. This is related to the zero-range
character of Skyrme forces, but it remains to a large extent true when finite-range 
pairing interactions are adopted \cite{Engel} --- also in the context of Relativistic Mean Field (RMF) 
calculations \cite{Paar03}. In particular, in Refs.~\cite{Engel,Bo,Yoshida1} it has been shown that 
in self-consistent HFB plus QRPA calculations the isoscalar pairing interaction shifts 
some low-energy Gamow-Teller (GT) strength
downwards, so that by fitting the n-p pairing strength (at least locally) one can account for the 
$\beta$-decay half-lives in neutron-rich nuclei.
The isoscalar pairing interaction is also important
for the double-$\beta$ decay~\cite{Su}.  
However, not only the isoscalar pairing
but also other terms of the effective interaction 
affect in an important way the main peak and low-energy part of the 
GT strength~\cite{SGII, Bender, Bai1, Bai2, Minato}: in particular,
this is true for the spin-orbit one-body potential, the spin two-body
terms and also tensor terms. One goal of Ref. \cite{Fr} was to
point out the necessity of improving Skyrme forces, and this goal
has been reached in Ref. \cite{Xavi12}. 

In short, although it has been put into evidence clearly that 
both the GTR and the higher order multipole of charge-exchange
transitions, such as the spin-dipole and spin-quadrupole transitions, 
will receive contribution from both the isoscalar and the isovector
pairing forces, firm constraints for the isoscalar pairing 
have not been extracted until very
recently. One of the reasons is that in many of the previous studies
the nuclei that have been considered possess neutron excess and 
are not close to the regions where
isoscalar pairing effects are expected to show up.
Thus, an important purpose of this contribution is to demonstrate that 
one is bound to 
consider specific nuclei and/or
specific properties to pin down 
unambiguous information about isoscalar pairing.


\section{Nuclear structure and isovector spin-singlet pairing interaction}
\label{Sect.02}
\subsection{Odd-even mass staggering}
\label{Sect.02-1}
\begin{figure}[ht]
\includegraphics[scale=.33,angle=0, bb=0 0 720 540]{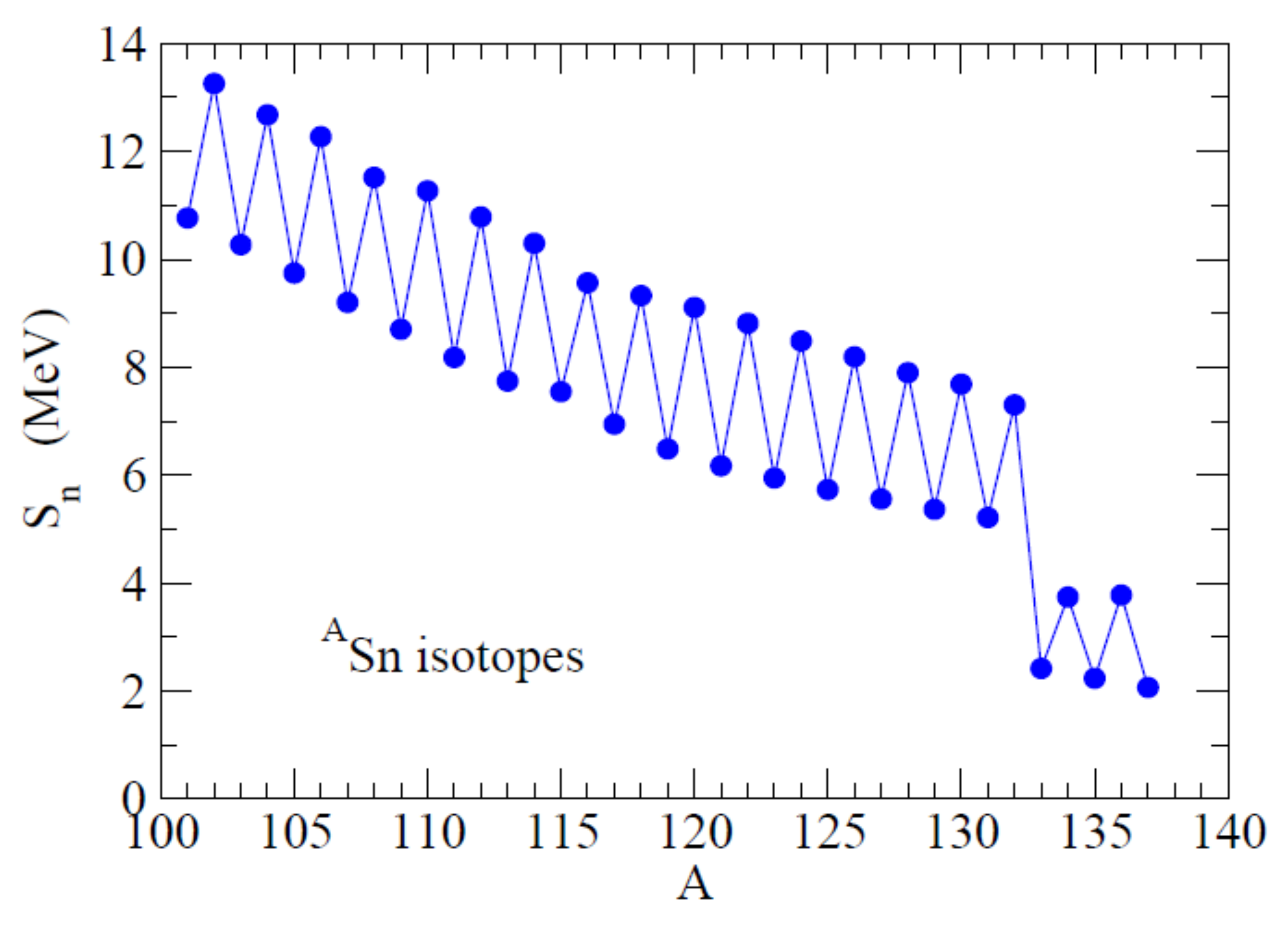}
\caption{\label{fig.01} Separation energy of Sn isotopes, as defined
in the text in Eq. (\ref{eq:sep_en}).} 
\end{figure}

The nuclear binding energies are found to show a systematic variation depending 
on the even or odd values of $Z$ and $N$,
\begin{equation}
\Delta \mathcal{B}= \left\{     \begin{array}{rl} 
     \Delta &  \rm{for}\,\,\, \it{Z}=\rm{even} \,\,\, and \,\,\,\it{N}\rm{=even}, \\
      0         &  \rm{for }\,\,\,\it{A}=\rm{odd},  \\
    -\Delta  &  \rm{for }\,\,\,\it{Z}=\rm{odd}\,\,\, and\,\,\, \it{N}\rm{=odd},
  \end{array}
  \right.  
\end{equation}
where $A=N+Z$. To illustrate this point, 
the separation energy which is the difference of binding energies 
$\mathcal{B}(A)$ of two neighboring nuclei, 
\begin{eqnarray}\label{eq:sep_en}
S_n(A)=\mathcal{B}(A)-\mathcal{B}(A-1),
\label{extra-B}
\end{eqnarray}
is shown in Fig. \ref{fig.01} in the case of Sn isotopes. The staggering 
of $S_n(A)$ is certainly due to the extra binding of even-$N$ Sn isotopes. 
Specific filters can be introduced to prove this evidence. 
In particular,
the 3-point formula for the neutron pairing gap, or pairing index, reads
\begin{eqnarray}\label{eq:3point}
\Delta^{(3)}(N)&=&(-)^{A+1}\frac{\mathcal{B}(N+1)-2\mathcal{B}(N)+\mathcal{B}(N-1)}{2} \nonumber  \\
&=&(-)^{A+1}\frac{S_n(A+1)-S_n(A)}{2}.
\end{eqnarray} 
This pairing index is expected to be proportional to the extra 
binding energy $\Delta$ in Eq. (\ref{extra-B}).
This is of course to be considered as a reasonable first approximation. Subtle
interferences between pairing effects and mean-field effects 
are discussed in Refs. \cite{Satula,Duguet,dug01}. 
The 3-point formula centered in the odd-$A$ nucleus can remove the major shell effect on 
the pairing index \cite{Satula}, while the 4- and 5-point formulas were also considered 
to avoid a large shell effect in the systematic studies of Refs. \cite{Duguet,dug01}. 
Again, to a first approximation, the pairing index can be
parametrized as
\begin{eqnarray}
\Delta^{(3)}(\rm{ A})\approx 12/{\rm A}^{1/2}\ MeV, 
\label{Delta-exp}
\end{eqnarray} 
in the broad region of the mass table 16 $<A<$ 250 \cite{Bohr1}.

As mentioned in the Introduction, HF-BCS or HFB are the standard
theories to describe pairing in nuclei. In such frameworks, the
quasi-particle energies are expressed as 
\begin{equation}\label{eq:BCSqp}
E_{k}=\sqrt{(\varepsilon_k-\lambda)^2+\Delta_k^2}, 
\end{equation}
where 
$\varepsilon_k$ and $\Delta_k$ are the single-particle energies and
(state-dependent) pairing gaps, respectively, and 
$\lambda$ is the chemical potential.
If we assume that the ground state of the odd nucleus is 
a quasi-particle state on top of the even core, then the comparison
between Eqs. (\ref{eq:3point}) and (\ref{eq:BCSqp}) clearly
shows that $E_k \approx \Delta_k \approx \Delta^{(3)}$.  

Only for n-n and p-p pairing such clear filters exist and
point to values of the pairing gap that are consistent with other
observables like those mentioned at the start of the Introduction.
From a theoretical viewpoint, however, the omission of n-p pairing has been
found to be not fully justified already fifty years ago (see 
\cite{Goodman1} and references therein). The complete 
generalized isospin pairing theory can be introduced 
\cite{Goodman1,Goodman2}. In the BCS version, the Cooper pairs are
not simply formed by two identical nucleons in time-reversed 
states, $\vert k \rangle$ and $\vert \tilde k \rangle$, but 
rather one can assume pairs made up in a general way with the four states 
$\vert k, \pi \rangle$, $\vert k, \nu \rangle$,
$\vert \tilde k, \pi \rangle$ and 
$\vert \tilde k, \nu \rangle$. 
In such generalized theory, nonetheless, the quasi-particle
energy keeps the same form as in (\ref{eq:BCSqp}), and
the pairing gap reads
\begin{equation}
\Delta_k ^2= \vert \Delta_{k\pi,k\nu} \vert^2 +
\vert \Delta_{k\pi,\tilde k\pi} \vert^2 + \vert
\Delta_{k\pi,\tilde k\nu} \vert^2.
\end{equation}
In HFB one can
even further generalize the wave function, by considering
pairs that do not correspond to time-reversed states. 

There is a subtle interplay between pairing and deformation: the first 
calculations \cite{Goodman:ADNP}
showed that different types of pairing arise if the nucleus is
spherical, axially symmetric or triaxial, respectively. Only
light nuclei with $N \approx Z$ are candidates for n-p pairing. Neutron
excess and/or large spin-orbit splittings hinder n-p
pairing as discussed in the Introduction.

One may wonder whether the results for the dominance of the
different kinds of pairing are confirmed within calculations that
employ realistic effective interactions. In the work of Ref.
\cite{Bertsch1}, extensive HFB calculations have been performed
with the hope of clarifying if and where $T=0$ pairing can
dominate over $T=1$ pairing. A simple yet realistic choice
has been adopted for the mean-field Hamiltonian, that is, a
Woods-Saxon potential plus spin-orbit. The pairing forces
have been chosen to be zero-range form, namely
\begin{equation}\label{eq:vol_pair}
V^{T=1}(\vec r_1,\vec r_2) = \hat{P}_sV_0 \delta\left(\vec r_1 - \vec r_2 \right),
\end{equation} 
and
\begin{equation}\label{eq:vol_pair2}
V^{T=0}(\vec r_1,\vec r_2) = \hat{P}_tfV_0 \delta\left(\vec r_1 - \vec r_2 \right), 
\end{equation} 
where $\hat{P}_s$ and $\hat{P}_t$ are the projectors onto the 
spin-singlet and spin-triplet channels, respectively: 
\begin{eqnarray}
\hat{P}_s=\frac{1}{4}-\frac{1}{4}\mathbf{\sigma}_p\cdot{\mathbf \sigma}_n,\ 
\hat{P}_t=\frac{3}{4}+\frac{1}{4}{\mathbf \sigma}_p\cdot{\mathbf \sigma}_n. 
\label{S-proj}
\end{eqnarray}
If these forces are adjusted so to reproduce at best the matrix elements of
realistic shell-model calculations, the ratio $f$ between the
$T=0$ and $T=1$ pairing strengths turns out to be $f \approx$ 1.6-1.7. It has been
found that, with the exception of few light nuclei, $T=0$
pairing cannot dominate the ground-state of stable nuclei despite its
larger strength. The reason is precisely the magnitude of the
spin-orbit splittings. In fact, if the nuclei are large
enough so that the spin-orbit effects are weakened, $T=0$
pairing can take over; such nuclei have
mass larger than $A\approx 130-140$ and are mostly
prone to be unbound via proton emission.

\subsection{Low-energy collective excitations}
\label{Sect.02-2}
\begin{figure}[ht]
\includegraphics[scale=.33,angle=0, bb=0 0 720 540]{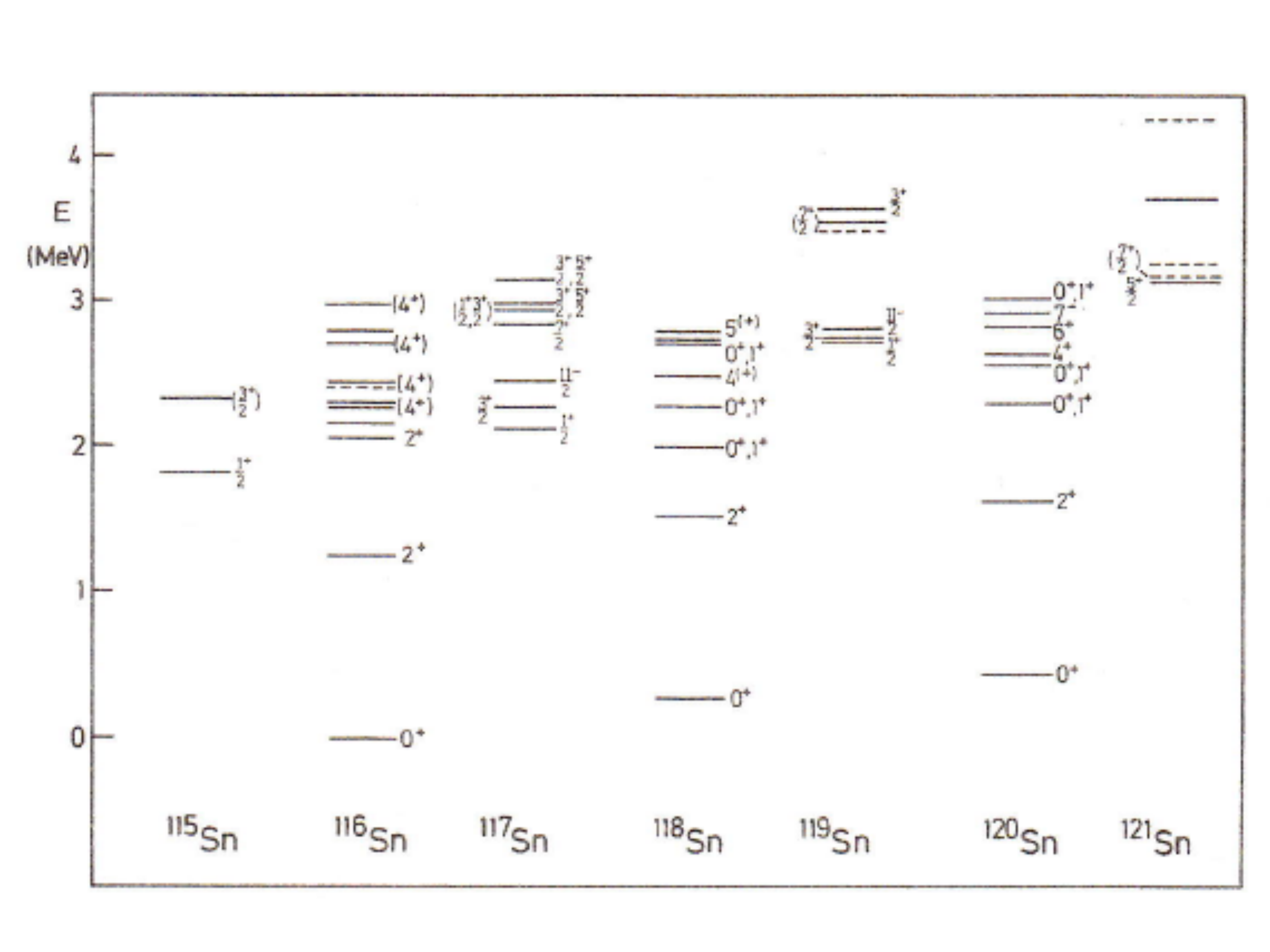}
\caption{\label{Sn-ex} Low-lying part of the energy spectra of 
few Sn isotopes. The figure is taken from Ref. \cite{Ring}. See the discussion
in the main text.}
\end{figure}

In Fig. \ref{Sn-ex} we display the low part of the excitation
spectra in few Sn isotopes. As is well known, and discussed above, 
in odd-$N$ isotopes the unpaired neutron does not feel pairing correlations
and can occupy several orbitals close to one another, so that there
is no gap in the excitation spectra. In even-$N$ isotopes, the 
lowest state is a 2$^+$ state. If we interpret this state as a
two quasi-particle (2qp) excited state, its excitation energy with
respect to the ground-state 0qp should be written as 
\begin{eqnarray}
E_{2qp}-E_{0qp} = 2E_{qp} \approx 2\Delta,
\end{eqnarray}
where the last approximation stems from the idea that both
excited quasi-particles lie at the Fermi surface. This latter 
equation has also a very transparent interpretation, namely it shows 
that the minimal energy to create an excited state of 2qp type
corresponds to breaking two Cooper pairs in order to recouple
the particles to different angular momenta.

In the tin region, the value of $\Delta$ is expected to be about 1.0 MeV 
[cf. Eq. (\ref{Delta-exp})]. 
However, the excitation energies of the $2^+$ states, as visible in 
Fig. \ref{Sn-ex}, are observed to be smaller than 2 MeV because of extra 
correlation energy.  
In fact, in QRPA, that is, the standard theory for such vibrational states, 
the energy of a collective state
$\hbar\omega$ can be schematically written as
\begin{equation}
\hbar\omega = \sqrt{ \left( E_{2qp}-E_{0qp} \right)^2 + \langle V 
\rangle^2},
\end{equation}
where $\langle V \rangle$ is an average value of the residual
interaction between quasi-particle states (cf. Ref. \cite{Ring} for
a complete account of QRPA with detailed formulas). In a magic
nucleus, in which pairing does not manifest itself, the residual
interaction would be only of p-h type, and the p-h channel
of the nuclear effective Hamiltonian is in a sense uniquely
sensitive to a given multipole component (e.g., quadrupole if
low-lying 2$^+$ states are under study). In open-shell nuclei
like the Sn isotopes, the p-p force plays an important role,
and low-lying states are sensitive both to its $J=0$ component
which is usually called pairing force and to its multipole
component. Last but not least, one should remind that not only
energies carry signatures of the p-p force as well as of other
components of the nuclear Hamiltonian, but electromagnetic 
transition probabilities $B(EL)$ do it as well \cite{Grodzins}.  Approximately, 
one could say that they are inversely proportional to energies.

\begin{figure}[ht]
\includegraphics[width=\columnwidth]{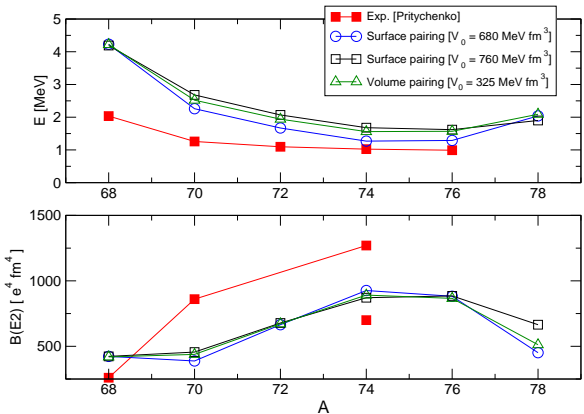}
\caption{\label{fig.ni_all} {Energies and electromagnetic transition
probabilities of the low-lying Ni isotopes calculated with different
pairing forces and compared with the experimental data.}}
\end{figure}

To disentangle such effects is not an easy task. Along the last decade,
fully microscopic QRPA calculations have become available
and have been extensively used to study low-lying vibrational
excitations. In Ref. \cite{Bertsch-2p}, for instance, the
global performance of QRPA with Skyrme forces has been
tested by calculating basically all nuclei that are experimentally
known. Unfortunately, only global conclusions have been extracted,
and the specific role of the pairing force has not been
pinned down as only one type of force has been used, namely
the so-called volume pairing force already defined in Eq. (\ref{eq:vol_pair}). 
This is the closest possible to a pairing
force with constant matrix elements $G$, since it is easy to understand that if
all wave functions have similar integral values within the nuclear
volume the radial matrix elements of such force cannot vary
much. Starting from the work \cite{Esbensen}, it has been
suggested that the pairing force should better be density dependent,
and may be taken as 
\begin{equation}\label{eq:T1surface}
V_{surface}^{T=1}(\vec r_1,\vec r_2) = \hat{P}_s V_0 \left( 1 - x\left( \frac{\rho}{\rho_0} 
\right)^{\gamma} \right) \delta\left(\vec r_1 - \vec r_2 \right), 
\end{equation}
with $x=1$ and $\gamma$ often taken as 1 for simplicity but
in principle arbitrary. Such an interaction is called surface pairing force, 
because if $\rho_0$ is around the saturation density the parameters
can be chosen so to enhance pairing at the surface. Several
studies have been shown that the low-lying states, and in particular
their transition probabilities, are very sensitive to the
choice of the pairing force (see, e.g., Ref. \cite{Carlsson}). Although some authors have expressed
preference for a surface pairing force in this context
\cite{Tolonnikov}, no conclusive study exists. We show in Fig. \ref{fig.ni_all} 
some results for the energy and reduced transition probability 
of the low-lying 2$^+$ states along the neutron-rich part of the Ni
isotope chain. The surface and volume pairing forces have been chosen so
that the values of the pairing gaps do not differ by more than 20\%. Only
in $^{78}$Ni the solution can be either supefluid or normal according
to the strength of the pairing force. This affects mainly the quadrupole
transition probability.

It would be of paramount importance to improve our understanding
of the effects of different type of pairing forces in the
low-lying nuclear spectroscopy. 3$^-$ or higher multipole states
can be of course discussed on equal footing. This could pave
the way to a better understanding of how pairing evolves far
from the stability valley, because the properties of low-lying 
vibrational states of even-even nuclei are among the very first 
observables that one can collect when producing new, exotic nuclei. 
At the same time, we stress that we are so far concerned only 
with the $T=1$ pairing force.
The microscopic origin of the density dependence of the pairing 
force, and/or the contribution to it coming from polarization effects, 
is ouside the scope of the present paper. 

\subsection{Moments of inertia of deformed nuclei}
\begin{figure}[tb]
\includegraphics[width=\columnwidth]{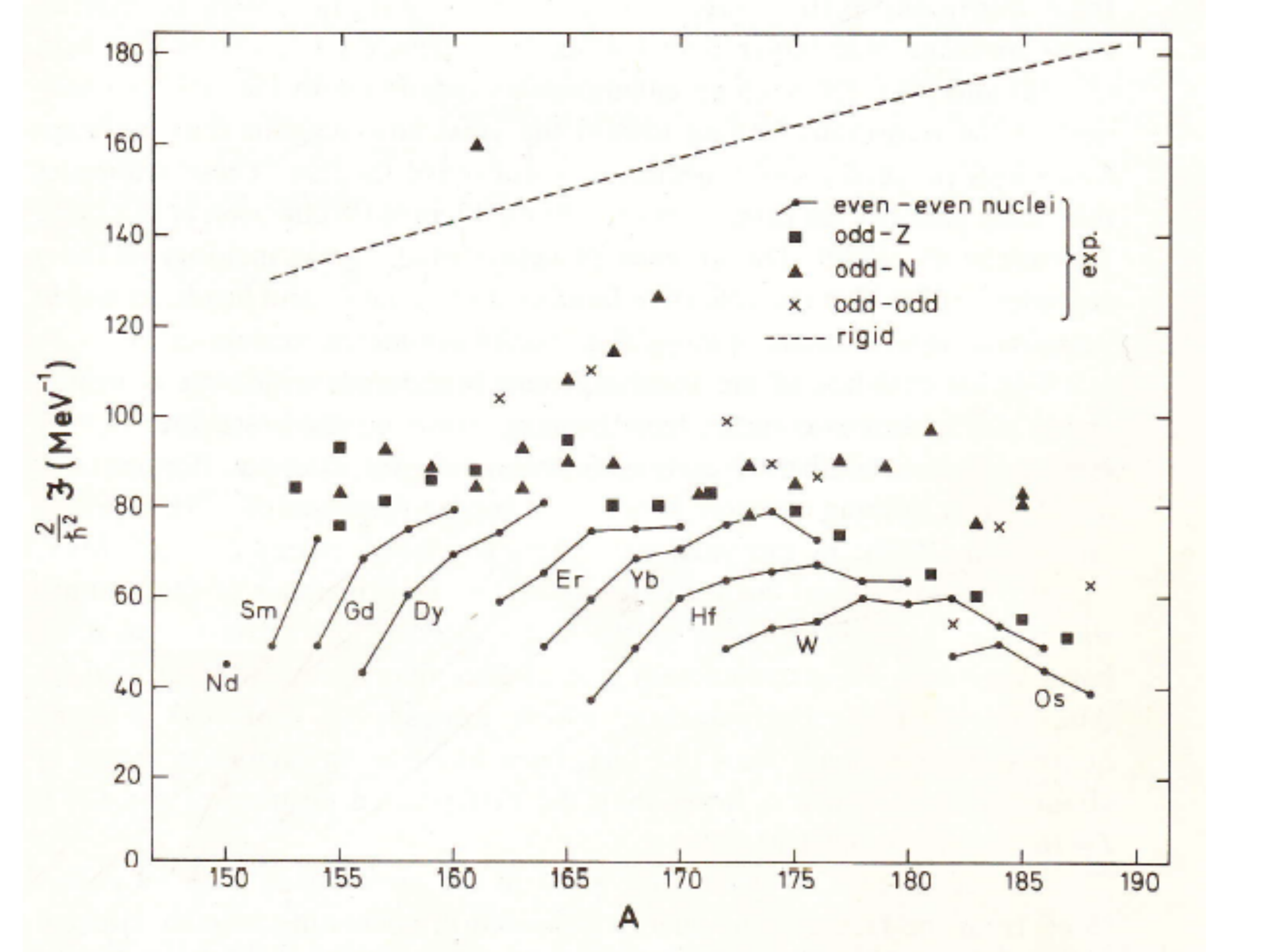}
\caption{\label{fig.03} Values of the moment of inertia of 
deformed nuclei with mass 150$<A<$188.  
The figure is taken from Ref. \cite{Bohr2}.}
\label{fig:moment-of-inertia}
\end{figure}

The energy spectrum of a deformed nucleus 
is characterized by the presence of rotational bands in
which levels with angular momentum $I$ have energies that scale
like
\begin{eqnarray}\label{eq:rotband}
E_I=E_0+\frac{I(I+1)}{2\Im}, 
\end{eqnarray}
where 
$E_0$ is the energy of the so-called bandhead, that is,
the level with a given intrinsic configuration and zero-point
rotational motion while
$\Im$ is the moment of inertia of the nucleus.
From the experimental point of view, the measurement of the
$\gamma$-transitions between the states (\ref{eq:rotband}) provides
the value of the energy of these states and consequently of the 
moment of inertia. From the theoretical point of view, it is possible
to perform so-called cranking calculations, in which the nuclear
Hamiltonian is solved with a constraint
$\omega j_x$ that corresponds to a rotation around the $x-$ axis
(we assume here that the nucleus has the $z-$ axis as a symmetry axis
and need to rotate around a perpendicular axis). In particular, 
one can obtain the value of the moment of inertia within the 
framework of second order perturbation theory (where the cranking term  
$\omega j_x$ plays the role of the perturbing field).   
The moment of inertia obtained in this way, so called Inglis's one, 
is given by
\begin{eqnarray}
\Im_{Inglis}=2\sum_{p,h}\frac{| \langle p \vert j_x \vert h \rangle 
|^2}{\varepsilon_p -\varepsilon_h},
\label{I-Inglis}
\end{eqnarray}
where the labels p and h corresponds to particle and hole states,
since the perturbing field creates p-h pairs at lowest order.
It is known that the Inglis moment of inertia is equivalent to the 
moment of inertia of a rigid body. Formally, this is proven 
within the harmonic oscillator potential approximation (cf. p. 77 of 
Ref. \cite{BM2}); however, the result can be empirically checked to
be the same in more realistic cases.  

In the presence of superfluidity, the ground state is represented by the BCS 
vacuum and the lowest order p-h excitations are replaced by 
the two quasi-particle ones. Accordingly, the moment of inertia 
can be obtained as
\begin{eqnarray}
\Im_{BCS}=2\sum_{\alpha,\beta}\frac{| \langle \alpha \vert j_x \vert 
\beta \rangle |^2(u_{\alpha}v_{\beta}-u_{\beta}v_{\alpha})^2}{E_{\alpha} +E_{\beta}}, 
\label{I-BCS}
\end{eqnarray}
where $E$ labels the quasi-particle energy [Eq. (\ref{eq:BCSqp})], and $u,v$ are the 
BCS unoccupation and occupation factors with 
$u_k^2+v_k^2=1$.
The formula (\ref{I-BCS}) gives in general smaller values than 
Eq. (\ref{I-Inglis}) since the denominator 
$E_{\alpha} +E_{\beta}$ is larger than $\varepsilon_i -\varepsilon_j$, 
and $(u_{\alpha}v_{\beta}-u_{\beta}v_{\alpha})^2$ is smaller than one.
The experimental values for the moment of inertia are compared with 
the calculated ones in Fig. \ref{fig:moment-of-inertia}.
The quenching with respect to the rigid-body value is one
of the evidences of superfluidity in nuclei.

\subsection{Pairing correlations at low density and BCS-BEC crossover}

\begin{figure}[ht]
\includegraphics[scale=.38,angle=0, bb=0 0 720 540]{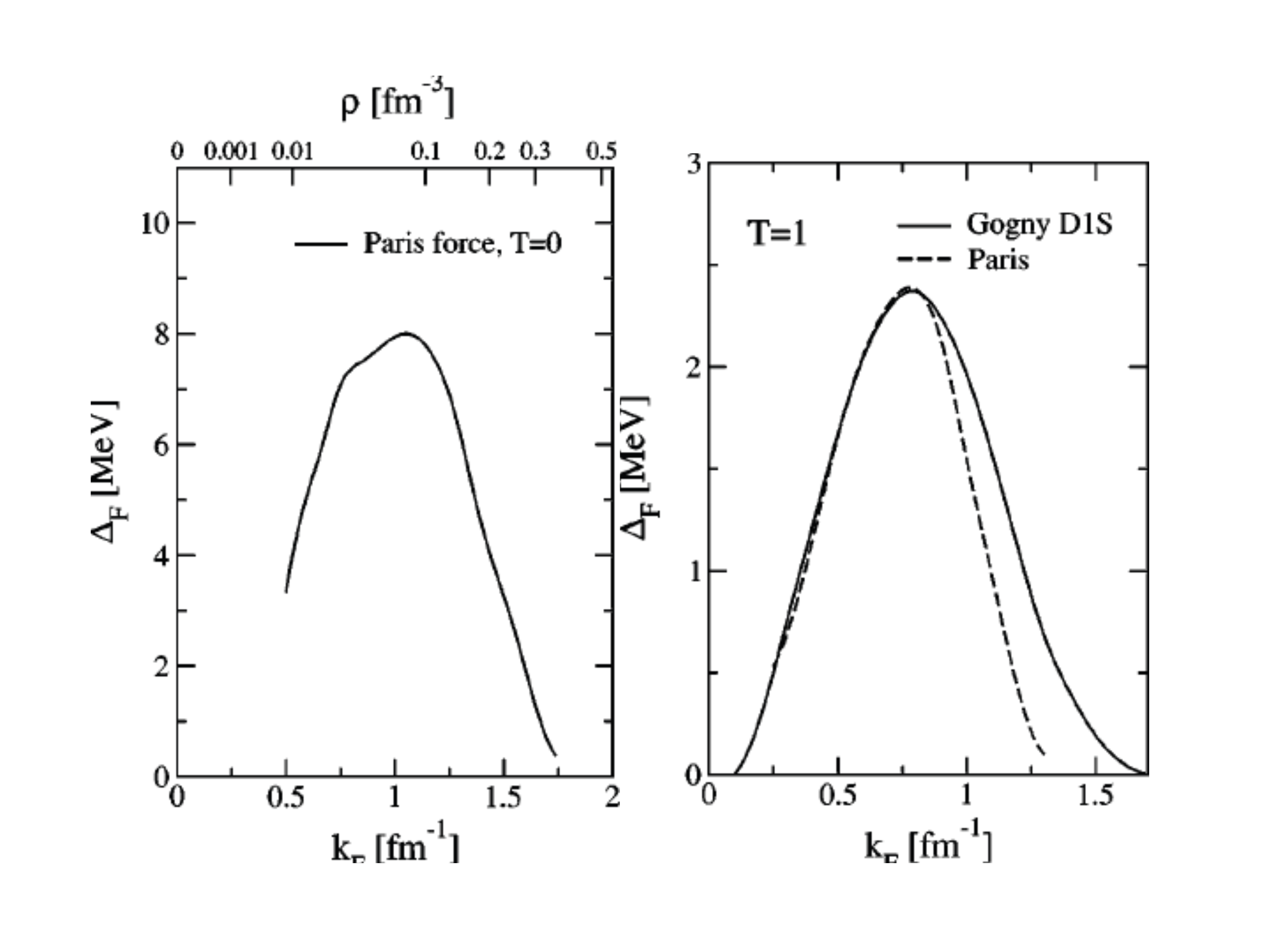}
\caption{\label{fig:gaps} Fermi momentum dependence of the pairing gaps, 
for the $T=0$ spin-triplet and the $T=1$ spin-singlet channels. 
The $T=0$ results are calculated by using the Paris potential, while the $T=1$ 
results are obtained by using either the Paris potential or the Gogny D1S effective force. 
The corresponding nuclear density is also given in the upper axis of the figure for 
the $T=0$ case. The figures are taken from Ref. \cite{Ga2}.} 
\end{figure}

\begin{figure}
\vspace{-4.2cm}
\includegraphics[scale=.45,angle=90, bb=0 0 720 540]{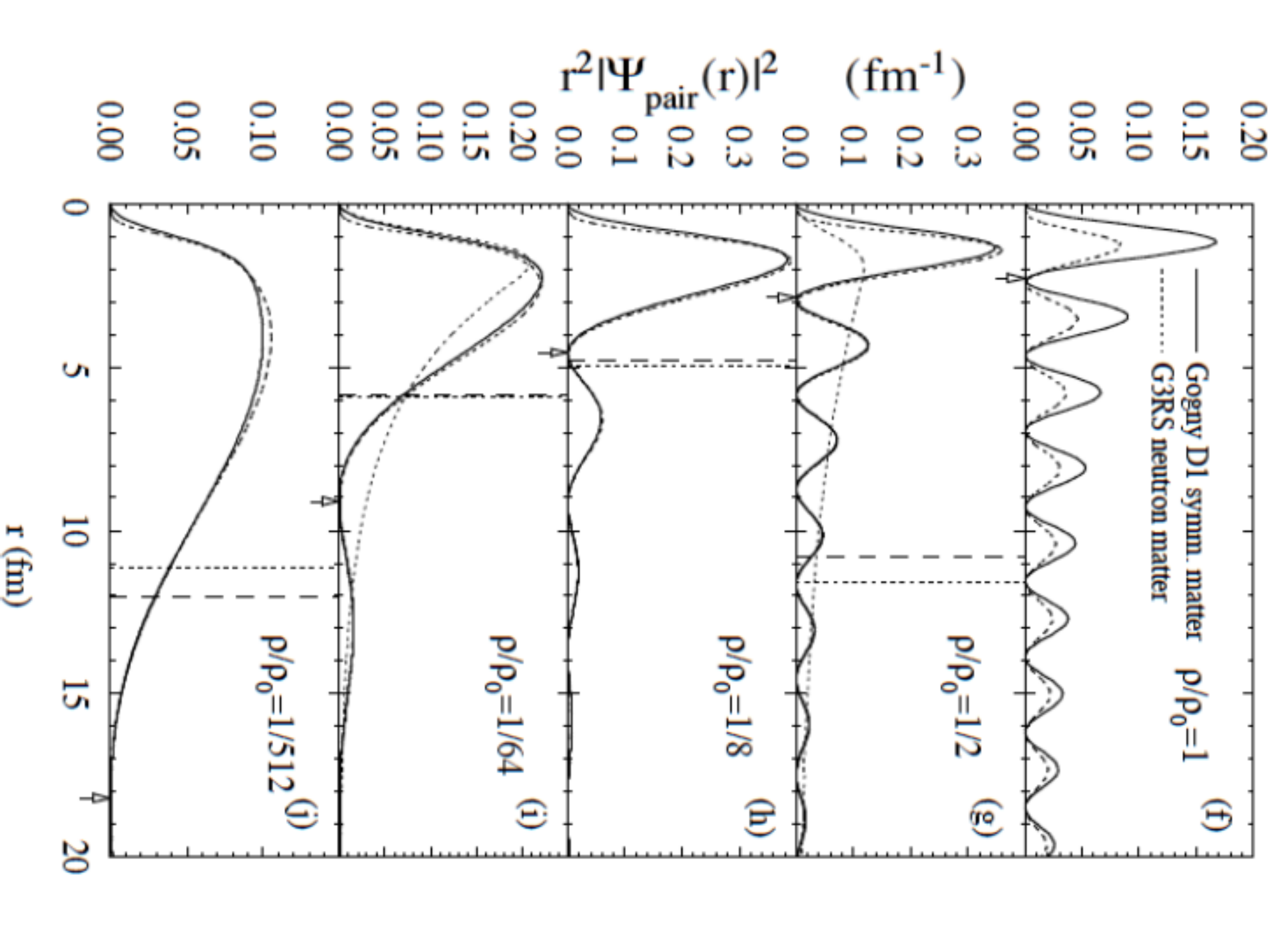}
\vspace{1cm}
\caption{\label{fig:BCS-BEC} The square of the wave function $\Psi_{pair}$ of the neutron 
Cooper pair, as a function of the relative distance $r$ between the pair
partners, at the neutron density $\rho/\rho_0$ = 1, 1/2, 1/8, 1/64, 1/512. 
The solid curve is for the pair in symmetric nuclear matter
calculated with the Gogny D1 force, while the dotted curve is for the pair in neutron 
matter calculated with the G3RS force (see also the main text). The vertical dotted
line represents the r.m.s. radius of the Cooper pair in neutron matter, 
while the dashed line represents the same quantity in the case of 
nuclear matter. The arrow indicates the average inter-neutron distance $d$. 
The thin dotted line in (g) and (i) is the wave function of a fictitious 
bound state in the free space (this is calculated by increasing the pairing strength, so
that a bound state appears, and tuning it so that the r.m.s. radius is the same of the 
real state whose wave function is displayed). The figure is taken from Ref. \cite{Matsuo} 
} 
\end{figure}

The pairing correlations at low nucleon density are of special interest, 
since the theoretical predictions for low-density uniform matter suggest that the 
pairing gap may take, at around 1/10
of the normal nuclear density, a value which is considerably larger than that 
around the normal density, 
both in the case of the isovector spin-singlet and the isoscalar spin-triplet channel
(see e.g. \cite{Ga2} and references therein).
This
feature is expected to have direct relevance for the properties of neutron stars, 
especially those associated with the inner
crust. The strong pairing at low density may also be relevant for finite nuclei, 
if one considers neutron-rich nuclei
near the drip-line as, for example, $^{6}$He and $^{11}$Li. This is because such nuclei 
are characterized by low-density distributions
of neutrons around the nuclear surface (the so-called neutron skin and/or neutron halo). 
It is interesting
to clarify to which extent the pair correlations in these exotic nuclei are different 
from those in stable nuclei, reflecting the strong
density dependence of the pairing correlations.  
In fact, 
the di-neutron correlation in the two-neutron halo nuclei such as $^{11}$Li, 
in which a spatially correlated pair formed by the halo neutrons shows up, 
may be considered as a manifestation of 
the strong pairing correlations in low-density regime.  
A recent theoretical analysis using the HFB method \cite{Pillet} also predicts 
the presence of similar di-neutron correlations in medium-mass
neutron-rich nuclei where more than two weakly-bound neutrons 
contribute to create the neutron skin in the outer part 
of the nuclear surface.

It has been argued that the BCS superconducting phase will change 
to a strong coupling regime, or Bose-Einstein condensate (BEC) phase 
made up with spatially compact, bound Fermion pairs if the pairing strength becomes 
strong enough to drive a phase transition.
The schematic solutions of the BCS equations \cite{Brink1} suggest that the 
relevant parameter in this respect is the pairing strength times the level density. 
In ordinary nuclei, the level density is as a rule too small to have a phase 
transition. Moreover, it is impossible to design systems where these parameters 
can be varied. Instead,
the BCS-BEC crossover
phenomenon was recently observed in ultra-cold atomic gases in a trap, for which the 
interaction is controllable \cite{Greiner}. In the case of the nuclear pairing, the 
BCS-BEC crossover has been argued mostly for the neutron-proton $T=0$ pairing in
the tensor coupled spin-triplet $^3SD_1$ channel, for which the strong 
deuteron-like spatial correlations 
may be the driving force, while at the same time uniform systems like matter
in neutron stars can realize a situation where the level density is large. 
Concerning the neutron pairing in the $T=1, S=0$ channel, we may expect strong 
coupling regimes in the low density matter, such as unique di-neutron or di-proton correlations,
whereas the transition to BEC phase is more questionable.

The pattern change of the spatial correlations of Cooper pairs in nuclear matter was 
discussed by M. Matsuo in Ref. \cite{Matsuo}. The square of the wave function of the 
neutron Cooper pair is shown, as a function of the relative 
distance $r$ between the two neutrons and at various densities, in Fig. \ref{fig:BCS-BEC}. 
The solid curves show the results in symmetric nuclear matter obtained with the Gogny D1 force, 
while the dotted curves are the results in neutron matter obtained with
G3RS force, which is a simple representation with three Gaussians of the bare
force \cite{ref:G3RS}.
At normal nuclear density, the wave function has an oscillatory behaviour in the coordinate space 
and 
the coherence length $\xi$ is large (the coherence length $\xi$ is the measure of the 
size of the Cooper pair, defined in a similar way to the r.m.s. radius of wave 
function \cite{Brink1}). This is a typical BCS-type wave function in the weak-coupling 
pairing scheme.  On the other hand, at lower density such as in the case in which 
$\rho/\rho_0$=1/8, the pairing correlations are strong enough to create a BEC-like 
wave function which has a short coherence length and 
is quite compact in coordinate space.

\begin{figure*}[htb]
\includegraphics[scale=.6,angle=0, bb=0 0 720 540]{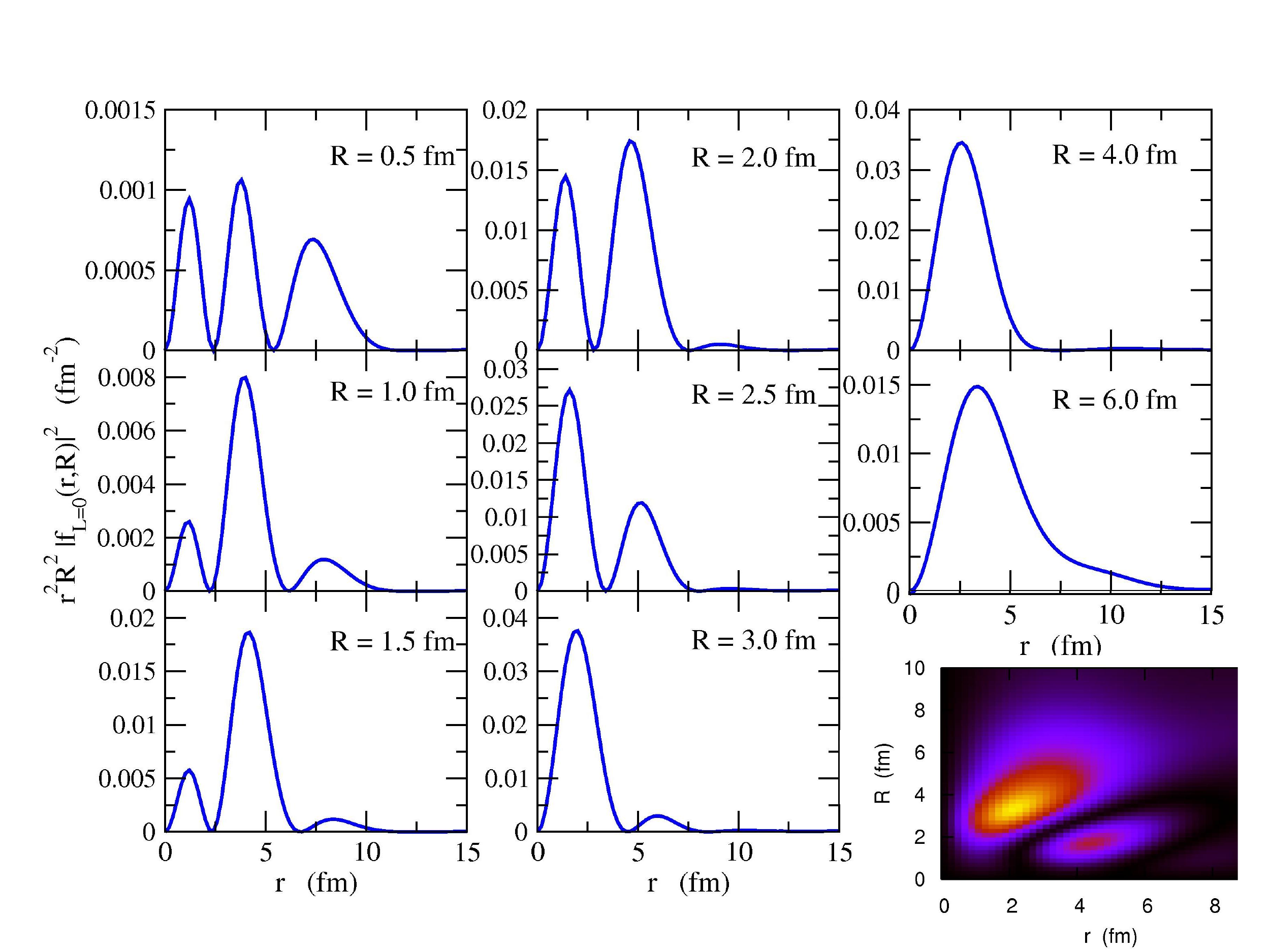}
\caption{\label{fig:Li11-pair} The square of wave function $f_{(L=0)}(r,R)$ of the neutron 
pair as a function of the relative distance $r$ between the neutrons. 
The bottom-right corner shows a 2D plot of the square of the wave function $f_{(L=0)}(r,R)$.  
The figure is taken from Ref. \cite{Hagino} 
} 
\end{figure*}

To move to finite, weakly-bound nuclei, 
we show in Fig. \ref{fig:Li11-pair} the neutron pair wave function in $^{11}$Li calculated by 
means of a
three-body model that assumes the $^9$Li core \cite{Esbensen,Bertsch}. 
The two-neutron wave function is calculated 
in the coordinate system where $\vec r_1$ and $\vec r_2$ are the distances of the
two neutrons from the center of the core. 
This wave function is then written in terms of the relative distance $\vec r$ between the 
two neutrons, and the
distance ${\vec R}$ between the core and the center of the two neutrons. The 
projected wave function on the total spin $S=0$ state reads
\begin{eqnarray}
&&\Psi^{(S=0)}(\vec r_1, \vec r_2) \nonumber \\
&=&\sum_{L}f_L(r,R)[Y_L({\hat r})
Y_L({\hat R})]^{(00)}[\chi_{1/2}\chi_{1/2}]^{(00)}.
\end{eqnarray}
The density dependent pairing interaction adopted in the three-body model calculations 
is strongly attractive near the surface, and rather weak near the center of nucleus, so that 
in terms of density dependence is quite similar to the realistic pairing 
interactions shown in Fig. \ref{fig:gaps}. The pair wave function in general has two peaks 
in the two-dimensional plane $(r,R)$, as 
shown in the bottom-right corner of Fig. \ref{fig:Li11-pair}. There is a strong peak at 
large $R$ and small $r$: this configuration is referred to as the "di-neutron" configuration. 
On the other hand, there is a smaller peak at large $r$ and small $R$: this is called 
the "butterfly" configuration.  
The total spin ($S$) component of the di-neutron configuration is dominantly $S=0$, 
while a large $S=1$ component is found in the butterfly configuration.  

The wave function associated with the di-neutron configuration is cut at different 
distances $R$ from the center and shown as well in Fig. \ref{fig:Li11-pair}. 
Near the center of the core at 
$R$ = 0.5 fm, the pair wave function has an oscillatory behavior similar to the nuclear matter 
result at the normal density $\rho/\rho_0$=1. When the value of $R$ is increased and the 
density becomes dilute, the oscillatory
behavior gradually disappears and one finds a single peak at at $R\approx$ 4 fm, where the pairing 
correlations reach their maximum. This di-neutron wave function in $^{11}$Li shows, therefore, 
a very similar behavior as the pair wave function in the nuclear matter 
that has been associated to the BCS-BEC crossover
and shown in Fig. \ref{fig:BCS-BEC}.   

\section{Competition between isoscalar and isovector pairing interaction}
\label{Sect.03}
\subsection{Pairing correlation energy for $pf$ shell configurations}
\label{Sect.03-0}

After having reviewed many general features of pairing in nuclei, we
aim in this Section to point to some observation that can pin down, 
quantitatively, the interplay between isovector and isoscalar pairing.
Certainly, pairing correlation energies are a quite direct observable: 
they represent the energy gain due to the pairing correlation.

We adopt in this case a separable form, namely the spin-singlet $T=1$ 
pairing interaction reads in this case
\be
V^{(T=1)}(\vec r_1,\vec r_2)=-G^{(T=1)}
\sum_{i,j}P^{(1,0)\dagger}_{i,i}(\vec r_1,\vec r_2)
P_{j,j}^{(1,0)}(\vec r_1,\vec r_2),
\label{eq:T=1}
\ee
where the pair field operator is defined as 
\be
P^{(T,S)\dagger}_{i,j}(\vec r_1,\vec r_2)=\delta_{l_i,l_j}\sqrt{2l_i+1}
[a^{\dagger}_{i}a^{\dagger}_{j}]^{(T,S)}\psi_i(\vec r_1)^*\psi_{j}(\vec r_2)^*,  
\ee
in terms of a single-particle wave function $\psi_i(\vec r)$ having 
quantum numbers $i\equiv\{n_i,l_i,j_i\}$. 
Here, $a_i^\dagger$ and $a_i$ are the creation and annihilation operators 
for the single-particle configuration $i$, respectively. 
The pairing strength $G^{(T=1)}$ can be fitted to the empirical pairing 
gaps given by Eq. (\ref{Delta-exp}) \cite{BM2,Brink1,Poves98}, 
and turns out to be in this case
\be
G^{(T=1)}=\frac{24}{A}~{\rm MeV}.
\label{eq:G-T1}
\ee
Even though the value 
in Eq. (\ref{eq:G-T1}) 
is a reasonable choice for calculations performed within a model
space consisting of one major shell
\cite{BM2,Brink1,Poves98}, 
the absolute value of the pairing strength 
should not be taken seriously since it depends on the model space adopted. 
It was pointed out in Ref. \cite{Mayo03} that the separable 
form of pairing interaction is 
quite useful as much as the non-separable realistic Hamiltonians 
adopted, e.g., in shell model calculations.

The spin-triplet $T$=0 pairing can also be given by a similar separable form,
\begin{eqnarray}
&&V^{(T=0)}(\vec r_1,\vec r_2)=\nonumber\\
&&-fG^{(T=1)}
\sum_{i\geq i',j\geq  j'}P^{(0,1)\dagger}_{i,i'}(\vec r_1,\vec r_2)
P_{j,j'}^{(0,1)}(\vec r_1,\vec r_2),\nonumber\\
\label{eq:T=0}
\end{eqnarray}
where the scaling factor $f$ 
is defined in close analogy as we have already done in Eqs. 
(\ref{eq:vol_pair}) and (\ref{eq:vol_pair2}), and as we shall do below. This factor 
is varied
between 1 and 2 in the following study of the pairing 
correlation energy \cite{Sagawa2013}. 

\begin{figure}[bt]
\hspace{-2.3cm}
\includegraphics[scale=0.4,clip,angle=-90.,bb=0 0 595 842]{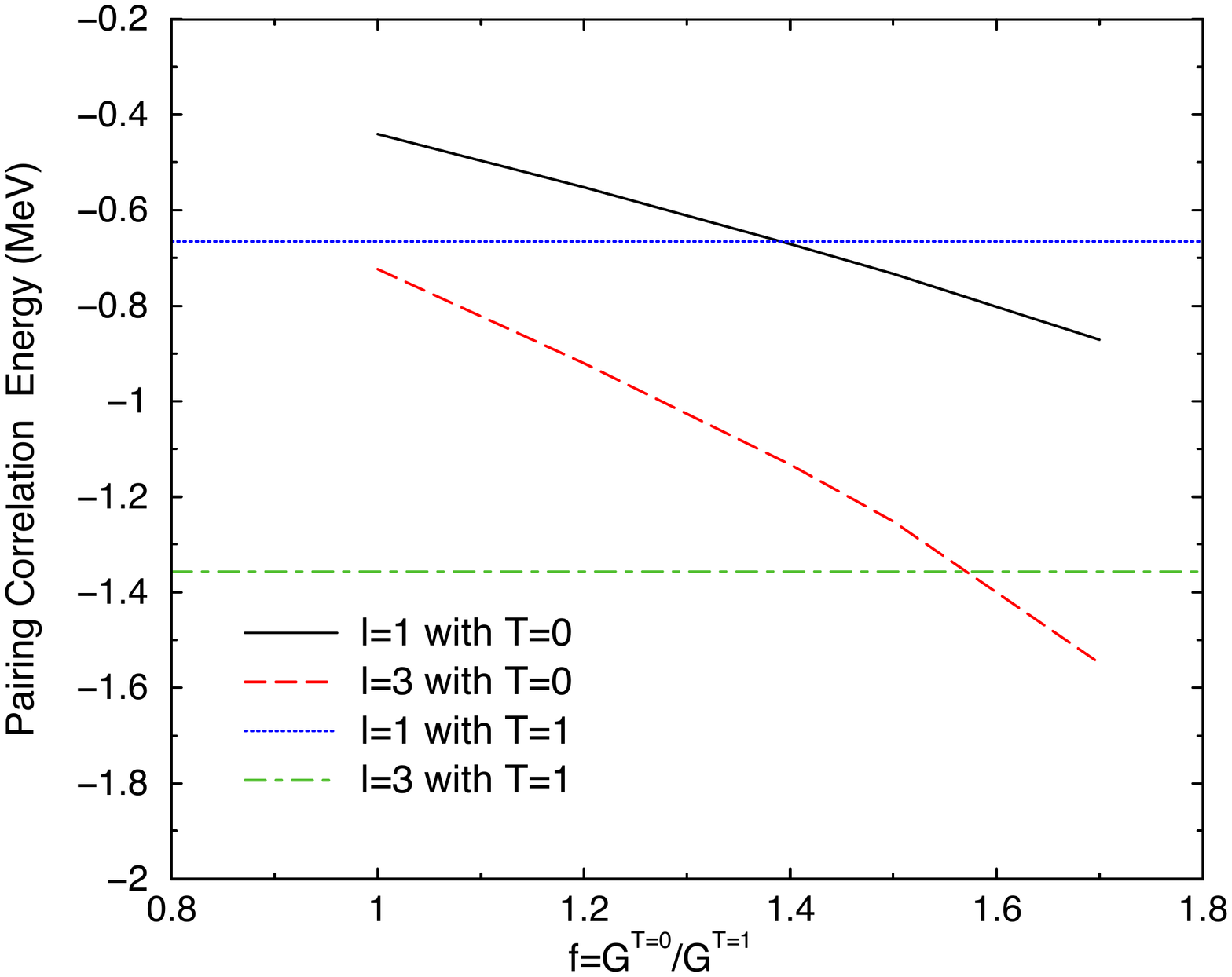}
\vspace{-1cm}
\caption{\label{T0T1pair}The pairing correlation energies for the 
lowest ($J^\pi=0^+$, $T$ = 1) 
and ($J=1^+$, $T$ = 0) states in the case of the $l=3$ 
and $l=1$ configurations, as a function of the scaling factor $f$ for 
the $T=0$ pairing, defined in Eq. (\ref{eq:T=0}). 
The strength of the 
spin-singlet $T$ = 1 pairing interaction is chosen to be 
$G^{(T=1)}$ = 24/$A$ MeV 
with fixed mass number $A$ = 56, while the strength for 
the spin-triplet $T$ = 0 pairing, $G^{(T=0)}$, 
is varied as it is equal to the factor $f$ 
multiplied by G$^{(T=1)}$. This figure is taken from Ref. \cite{Sagawa2013}.}
\end{figure}

We can now discuss the energy gain due to the pairing correlations, 
that is, the pairing correlation energy.  
Fig. \ref{T0T1pair} shows these energies
for the $p$-orbit ($l=1$) and the $f$-orbit ($l=3$) configurations, 
as a function of the scaling factor $f$ for the $T=0$ 
pairing. The energies for both 
the $J^{\pi}=0^+$ state with the isospin $T=1$, and the $J^{\pi}=1^+$ state 
with the isospin $T=0$ are shown in the figure. 
In the case of the $T=0$ pairing, we need to add to the Hamiltonian
the spin-orbit splitting parametrized as 
\be \label{s-o}
\Delta \varepsilon_{ls} =-V_{ls}({\vec l}\cdot{\vec s}), 
\ee
where the spin-orbit coupling strength $V_{ls}$ is taken to be \cite{BM2}
\be \label{s-o-c}
V_{ls}=\frac{24}{A^{2/3}}~{\rm MeV}.
\ee
This spin-orbit potential reproduces well the empirical 
spin-orbit splitting $\Delta\varepsilon$ = 7.0 MeV 
between the $1f_{7/2}$ and $1f_{5/2}$ states in $^{41}$Ca \cite{Uozumi}.

We diagonalize the pairing Hamiltonian 
separately for the $p$- and $f$-orbit 
configurations in order 
to disentangle the roles of the pairing and of the spin-orbit interactions 
in a transparent way. It should be noticed that, for the $T=0$ pairing, the pair configurations 
are constructed not only with two equal orbitals having 
$(l_i=l_{i'},j_i=j_{i'})$ but also with the spin-orbit partner orbits 
{\em viz.} $(l_i=l_{i'},j_i=j_{i'}\pm 1)$, as is seen in the two-body $T=0$ pairing 
matrix elements discussed in the Appendix [cf. 
Eq. (\ref{eq:T=0me})]. Thus,  
in the $l=1$ case, the ($2p_{3/2})^2$ and ($2p_{1/2})^2$ configurations
are available for the $T=1, J^{\pi}=0^+$ state, while also the ($2p_{3/2}2p_{1/2})$ 
configuration is available for the $T=0, J^{\pi}=1^+$ state. 
In a similar way, the 
$(1f_{7/2})^2$ and $(1f_{5/2})^2$ 
configurations participate to the $J^{\pi}=0^+$ state in the $l=3$ case,  
and also the ($1f_{7/2}1f_{5/2})$ configuration is involved in the 
$J^{\pi}=1^+$ state.  

As one can see in Fig. \ref{T0T1pair}, 
the lowest energy state with $J^{\pi}$ = 0$^+$ 
for the $l=3$ case gains more binding energy than 
the $J^{\pi}$=1$^+$ state if the factor $f$ is smaller than 1.5.  
In the strong $T$ = 0 pairing case, 
that is, $f\geq 1.6$, 
the $J^{\pi}$=1$^+$ state obtains more binding energy 
than the lowest $J^{\pi}$ = 0$^+$ state.
These results are largely due to the quenching of the $T$ = 0 
pairing matrix element
by the transformation coefficient from the $jj$ to $LS$ coupling schemes 
\cite{Sagawa2013}. 
This quenching never happens for the $T$=1 pairing matrix element, 
since the mapping of the two-particle wave function 
between the two coupling schemes 
is simply implemented by a factor $\sqrt{j+1/2}$ in Eq. (\ref{eq:T=1me}).  
In the $l=1$ case, the competition between the $J^{\pi}$ = 0$^+$ 
and the $J^{\pi}$ = 1$^+$ states
is also seen in Fig. \ref{T0T1pair}.  
Because of the smaller spin-orbit splitting in this case,  
the couplings among the available configurations are rather strong, 
and the lowest $J^{\pi}$ = 1$^+$ state gains more binding energy 
than the $J^{\pi}$ = 0$^+$ state in the case of $f\geq 1.4$. 
These results are consistent with the observed spins 
of $N=Z$ odd-odd nuclei in the $pf$ shell, 
where all the ground states have the spin-parity $J^{\pi} = 0^+$, 
except for $^{58}_{29}$Cu.
The ground state of $^{58}_{29}$Cu has $J^{\pi} = 1^+$, 
since the odd proton and odd neutron occupy mainly 
the $2p$ orbits, where the spin-orbit splitting is expected to be much 
smaller than that of $1f$ orbits.
 
The mass number dependence of the spin-orbit splitting is approximately 
determined by Eq. (\ref{s-o-c}). 
Since the strength of the spin-orbit potential 
and the largest angular momentum 
in each major shell are proportional to 
$A^{-2/3}$ and $A^{1/3}$ \cite{Bohr2}, respectively, 
the spin-orbit splitting of the largest angular momentum states 
is roughly proportional to $A^{-1/3}$.
On the other hand, the $T=0$ pairing matrix elements (\ref{eq:T=0me}) for 
the separable interaction are proportional to the product of the pairing 
strength $G^{(T=0)}$ and the angular momentum of the valence orbit. This $T=0$ pairing 
matrix element is thus expected to be proportional 
to $A^{-1}\times A^{1/3}=A^{-2/3}$. 
Thus, the spin-orbit splitting decreases more slowly than the 
pairing matrix element as a function of 
the mass number $A$.  
As a result, 
the spin-orbit splitting 
hinders more effectively the spin-triplet pairing correlations 
in medium-heavy nuclei with $N=Z>30$ 
compared with lighter nuclei with $N=Z<30$.  
We should also mention that, in reality, 
the spin-orbit splitting decreases even 
more slowly than predicted by the approximate $A^{-1/3}$ dependence:
that is, it is found to be 
6.2 MeV for the $l=1$ states in $^{16}$O, 5.5 MeV for the $l=2$ states 
in $^{40}$Ca, 7.0 MeV for the $l=3$ states in $^{56}$Ni, 
and 7.0 MeV for the $l=4$ states in $^{100}$Sn \cite{Bohr2,Xavi12}. 

As mentioned above,
the shell model matrix elements are consistent with a factor $f$ 
in Eq. (\ref{eq:T=0}) in the range of 1.6-1.7, for both $sd$ shell and
$pf$ shell configurations \cite{BL10,Brown06,Honma04}.  
In Ref. \cite{Poves98}, the ratio 1.5 is adopted to 
analyze the spin-triplet pairing correlations 
in the $N=Z$ nuclei within shell model calculations.
These adopted values of $f$, together with the results 
shown in Fig. \ref{T0T1pair}, 
suggest that, in the odd-odd 
$N=Z$ nuclei, the configuration with $J^{\pi}=1^+$ is favored in 
the ground state rather than the $J^{\pi}=0^+$ one, especially  
when the $p_{3/2}$ orbit is the main configuration for the 
valence particles.  
However the onset of spin-triplet pair condensation is not be
guaranteed by the simple inspection of the spin of the ground state, and may 
need a careful examination of many-body wave 
functions emerging from HFB or large-scale shell model calculations  
\cite{Gezelis11}. 

\subsection{IAR and Gamow-Teller states in normal nuclei}
\label{Sect.03-1}

\subsubsection{RPA model and formalism}
\label{Sect.03-1a}

We recall here the main features of charge-exchange RPA and 
QRPA in the standard matrix formulation. The elementary excitations are either proton 
particle-neutron hole pairs of the type 
$a^{\dagger}_{\alpha,\pi} a_{\beta,\nu}$, or neutron particle-proton 
hole pairs of the type $a^{\dagger}_{\alpha,\nu} a_{\beta,\pi}$.
Without pairing correlations, i.e., when the ground-state
is HF, the index
$\alpha$ ($\beta$) labels in what follows the unoccupied 
(occupied) states. Within the linear response theory, 
or RPA, the excited states $\vert n,\pm \rangle$ 
result from the action of the operators $\Gamma^{\dagger}_{n,\pm}$ on 
the correlated RPA ground state denoted by $\vert\tilde 0\rangle$ for a nucleus with 
proton number $Z$. The RPA states have well-defined $\Delta T_z$, and 
here the label $\pm$ distinguishes the $T_-$ and $T_+$ 
modes leading respectively to the $Z+1$ and $Z-1$ nuclei.
The RPA operators can be expressed as follows:
\begin{eqnarray}\label{eq:RPAop}
\Gamma^{\dagger}_{n,-}&=&\sum_{\alpha,  \beta} 
X^{(n)}_{\alpha, \nu; \beta, \pi} a^{\dagger}_{\alpha, \pi} a_{\beta, \nu} 
- \sum_{\alpha, \beta}
Y^{(n)}_{\alpha, \nu;\beta, \pi} a^{\dagger}_{\beta, \pi} a_{\alpha, \nu},  
\nonumber \\
\Gamma^{\dagger}_{n,+}&=&\sum_{\alpha, \beta} 
X^{(n)}_{\alpha, \nu;\beta, \pi} a^{\dagger}_{\alpha, \nu} a_{\beta, \pi} 
- \sum_{\alpha, \beta}
Y^{(n)}_{\alpha, \pi;\beta, \nu} a^{\dagger}_{\beta, \nu} a_{\alpha, \pi}. 
\nonumber \\
\end{eqnarray}
With the additional assumption that the correlated RPA 
ground state is the vacuum for the RPA operators 
$\Gamma_{n,\pm}$ one can show that the $X^{(n)}, 
Y^{(n)}$ amplitudes define the eigenvectors of the following 
RPA secular matrix (where the indices $\alpha$ and 
$\beta$ have been dropped for simplicity): 
\begin{eqnarray}\label{eq:matrixRPA}
&& \left( \begin{array}{cccc}
A_{\pi\nu,\pi'\nu'} & 0 & 0 & B_{\pi\nu,\nu'\pi'} \\
0 & A_{\nu\pi,\nu'\pi'} & B_{\nu\pi,\pi'\nu'} & 0 \\
0 & -B_{\pi\nu,\nu'\pi'} & -A_{\pi\nu,\pi'\nu'} & 0 \\
-B_{\nu\pi,\pi'\nu'} & 0 & 0 & -A_{\nu\pi,\nu'\pi'} \\
\end{array} \right) \times \nonumber \\
&& \times
\left( \begin{array}{c}
X^{(n)}_{\pi'\nu'} \\
X^{(n)}_{\nu'\pi'} \\
Y^{(n)}_{\pi'\nu'} \\
Y^{(n)}_{\nu'\pi'} 
\end{array} \right) = E_{n} 
\left( \begin{array}{c}
X^{(n)}_{\pi\nu} \\
X^{(n)}_{\nu\pi} \\
Y^{(n)}_{\pi\nu} \\
Y^{(n)}_{\nu\pi} 
\end{array} \right)~.
\end{eqnarray}
The corresponding eigenvalues $E_n$ are the excitation energies of 
the RPA modes. The expressions of the $A$ and $B$ matrices are
\begin{eqnarray}
A_{\pi\nu,\pi'\nu'} & \equiv & A_{\alpha \pi,\beta \nu;
\alpha' \pi',\beta',\nu'} \nonumber \\
& = & \left( \varepsilon_{\alpha} - \varepsilon_{\beta} \right) 
\delta_{\alpha\alpha'} \delta_{\beta\beta'} + \langle \alpha \beta' 
\vert V_{\rm ph} \vert \beta \alpha' \rangle,
\nonumber \\
B_{\pi\nu,\nu'\pi'} & \equiv & B_{\alpha \pi,\beta \nu;
\alpha' \nu', 
\beta' \pi'} = \langle \alpha \alpha' \vert V_{\rm ph} \vert 
\beta \beta' \rangle,
\label{eq:ab_matrix}
\end{eqnarray}
and similarly for the other cases. 
In these formulas the single-particle energies $\varepsilon$ 
and the matrix elements of the residual particle-hole (p-h) interaction 
$V_{\rm ph}$ appear. 
At variance with the case of normal RPA, 
the $A$ matrix breaks into two blocks having different
dimensions, and the matrix $B$ is made up with two rectangular
blocks. 
Under the spherical symmetry 
assumption the RPA excited states have good angular momentum 
and parity $J^\pi$ and therefore, each $J^\pi$-mode corresponds 
to a separate diagonalization in the corresponding $J^\pi$ 
p-h space. In this case, all previous
expressions can be cast in angular momentum coupled form: the
final result is that the matrix equation (\ref{eq:matrixRPA})
retains its structure and the matrix elements are coupled in this way:
in
$\langle \alpha \beta' 
\vert V_{\rm ph} \vert \beta \alpha' \rangle$ the coupling is between
$\left( \alpha \otimes \tilde \beta \right)_J$ and similarly
$\left( \alpha' \otimes \tilde \beta' \right)_J$.
 
In open-shell nuclei pairing correlations must be taken into account
and, within the HFB framework the independent particles are replaced by
quasi-particles so that RPA becomes QRPA. 
Quasi-particles are mixtures of particles and holes. 
If we use the symbol $\alpha^{\dagger}$ ($\alpha$) for the quasi-particle 
creation (annihilation) operator, the QRPA modes are generated by
\begin{equation}\label{eq:QRPAop}
\Gamma^{\dagger}_{n} = \sum_{\alpha, \pi; \beta, \nu}
X^{(n)}_{\alpha, \pi;\beta, \nu} \alpha^{\dagger}_{\alpha, \pi} 
\alpha^{\dagger}_{\beta, \nu} - Y^{(n)}_{\alpha, \pi;\beta, \nu} 
\alpha_{\beta, \nu} \alpha_{\alpha, \pi}. 
\end{equation}
Also this formula, and the following ones, 
can be recast in angular momentum coupled form.
Similarly as in RPA, in QRPA also  the states have well-defined 
$\Delta T_z$. The QRPA matrix equation has the form 
\begin{equation}\label{eq:matrixQRPA}
\left( \begin{array}{cc}
A_{\pi\nu,\pi'\nu'} & B_{\pi\nu,\pi'\nu'} \\
-B_{\pi\nu,\pi'\nu'} & -A_{\pi\nu,\pi'\nu'} \\
\end{array} \right)
\left( \begin{array}{c}
X^{(n)}_{\pi'\nu'} \\
Y^{(n)}_{\pi'\nu'}
\end{array} \right) = E_{n}
\left( \begin{array}{c}
X^{(n)}_{\pi\nu} \\
Y^{(n)}_{\pi\nu}
\end{array} \right)~, 
\end{equation}
where once more the indices $\alpha$ and $\beta$ have been dropped.
The matrix elements appearing in the last formula include
both the particle-hole and the particle-particle (p-p) residual
interaction. QRPA can be built on top of HFB, or
on top of the simpler HF-BCS approximation. The QRPA
matrix elements display a quite similar form in the HF-BCS case
and in the HFB case, provided one uses the canonical basis.
They read
\begin{eqnarray}
A_{\pi\nu,\pi'\nu'} 
& = & \left( E_{\alpha,\alpha'}\delta_{\beta\beta'} 
+ E_{\beta,\beta'}\delta_{\alpha\alpha'} \right) \nonumber \\ 
& + & \left(u_\alpha v_\beta u_{\alpha'} v_{\beta'} +
v_\alpha u_\beta v_{\alpha'} u_{\beta'} \right)  
\langle \alpha \beta' \vert V_{\rm ph} \vert \beta \alpha' \rangle
\nonumber \\
& + & \left(u_\alpha u_\beta u_{\alpha'} u_{\beta'} +      
v_\alpha v_\beta v_{\alpha'} v_{\beta'} \right) 
\langle \alpha \beta \vert V_{\rm pp} 
\vert \alpha' \beta' \rangle,
\nonumber \\
B_{\pi\nu,\pi'\nu'} 
& = & \left(u_\alpha v_\beta v_{\alpha'} u_{\beta'} +      
v_\alpha u_\beta u_{\alpha'} v_{\beta'} \right)
\langle \alpha
\alpha' \vert V_{\rm ph} \vert \beta \beta' \rangle
\nonumber \\
& - & \left(u_\alpha u_\beta v_{\alpha'} v_{\beta'} +
v_\alpha v_\beta u_{\alpha'} u_{\beta'} \right)
\langle \alpha \beta \vert V_{\rm pp} 
\vert \alpha' \beta' \rangle. \nonumber \\
\label{eq:ab_qmatrix}
\end{eqnarray}
Here $u$ and $v$ are the usual unoccupation and occupation factors, respectively, 
either of the BCS or of the canonical states. $E_{\alpha\alpha'}$ is 
either $E_\alpha \delta_{\alpha\alpha'}$ (being $E_\alpha$ the BCS 
quasi-particle energy), or the canonical basis matrix element of
the HFB Hamiltonian. 

Having the (Q)RPA amplitudes $X^{(n)}$ and $Y^{(n)}$ at 
one's disposal, it is possible to calculate the transition
probability of the state $\vert \nu \rangle$ associated with a given 
operator $\hat O$.  It is clear that a state characterized
by many p-h or two quasi-particle configurations that do contribute 
coherently in the sum, possesses a large transition strength. 
The strength function is defined as
\begin{equation}
S(E) = \sum_\nu \vert \langle n \vert\vert {\cal O}_{\pm} 
\vert\vert \tilde 0 \rangle \vert^2 \delta(E-E_n), 
\end{equation}
and its moments are
\begin{equation}
m_k = \int dE\ E^k S(E) = \sum_\nu \vert \langle n \vert\vert
{\cal O}_{\pm} \vert\vert \tilde 0 \rangle \vert^2 E_n^k.
\end{equation}

Note that quite different ingredients are used even in self-consistent
RPA or QRPA by different groups. In the case of Skyrme calculations,
the spin-isospin component of the residual interaction in the p-h channel
is supplemented by a zero-range p-p force. In the case of RMF calculations,
in the residual interaction the pion exchange is introduced; since this
has no contribution for the ground-state which is treated at the Hartree
level, the corresponding coupling constant must be fitted. In the case
of RHF, the pion is introduced and fitted at the ground-state level: 
however, its role is modest in the residual interaction, and the position
of the change-exchange states turn out to be mainly determined by the
exchange terms associated with the isoscalar $\sigma$ and $\omega$ mesons \cite{Liang}.
Pairing is usually treated non-relativistically, both in RMF and RHF. 
This is important to be kept in mind, because even if our focus is on
pairing effects, they cannot be completely decoupled from the underlying mean
field.

\subsubsection{Results}
\label{Sect.03-1b}

As we have mentioned already, we expect that the IAR is sensitive,
in open-shell nuclei, to only or mainly $T=1$ pairing. In the
previous Section it is clearly shown that the p-n matrix elements
enter in the QRPA residual interaction. It could be possible 
to just fit these matrix elements. Although, generally, one
adopts pairing strengths of the order of what predicted by 
Eq. (\ref{eq:G-T1}), the p-p and n-n empirical
matrix elements have been often taken with different
strengths: for instance, the values
$G_n = 21/A$ and $G_p = 26/A$ have been adopted in Ref. 
\cite{Ragnarsson}. However, the Coulomb (anti-)pairing effect should
be taken into account and yet few groups have done it 
intensively \cite{Anguiano}. This effect may reduce the pairing gaps
by 100-200 keV \cite{Schuck-Coul-pair} which means between 
about 5\% and 20\%. Therefore, it is legitimate to assume that
the pairing force is isospin invariant, as all other components 
of any bare and/or
effective nuclear forces are. In this case, one
can say that ground-state 0$^+$ p-p and n-n matrix elements 
are equal to the p-n ones. 

\begin{figure}[ht]
\includegraphics[scale=0.35]{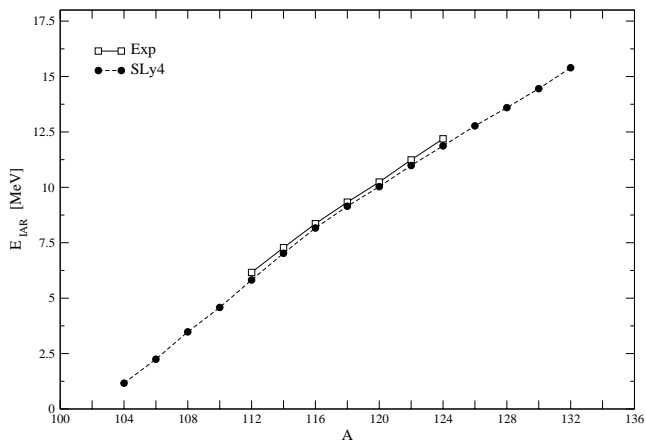}
\caption{\label{fig.Eiar} Energies of the IAR in the Sn isotopes
obtained by Skyrme-QRPA calculations are compared with experimental data 
from Ref. \cite{Pham}. The figure is taken from 
Ref. \cite{Fracasso-IAR}.}
\end{figure}

At least, this is the attitude that has been adopted in the 
context of both Skyrme \cite{Fracasso-IAR} and RMF \cite{RMF-IAR}
calculations of the IAR, in which the $T = 1$ residual p-n matrix
elements have been calculated consistently with the ground-state
pairing force and the assumption of isospin invariance. The results
for the IAR energies agrees very well with experiment (cf. 
Fig. \ref{fig.Eiar}), and it has
been shown, moreover, that without p-n pairing in the residual
interaction the IAR is fragmented in more than one peak and
does not exhaust the whole $N-Z$ sum rule as it should.

The Gamow-Teller resonance is instead sensitive to the largely
unknown p-n $T = 0$ pairing. Unfortunately, in standard nuclei this
sensitivity is not strong enough to allow to pin down the strength
of such pairing force. In Refs. \cite{RMF-IAR} it has been
found that the energy of the main GT peak in $^{118}$Sn 
changes by at most 100 keV when the $T = 0$ pairing strength
is changed by 50\%. The inclusion of $T = 0$ 
pairing reduces the configuration splitting in the region
of the main GTR, and has more influence generally speaking
on the low-lying strength as it can be expected. In fact,
it could be said that $T = 0$ pairing does play a role
in the GT excitation spectra because of the partial occupations
that are, in turn, induced by the $T = 1$ pairing in the ground state (this latter
being, of course, more active around the Fermi surface).

As already discussed, the GT strength function is more
sensitive to the spin-orbit splitting and p-h force.
Nonetheless, it is interesting to notice that despite
the rather different ansatz for such ingredients in
the RMF and Skyrme frameworks, some of the conclusions
reached in \cite{Fr} are the same as in \cite{RMF-IAR}.
In particular, even in the Skyrme calculations of
\cite{Fr} the reduction of the configuration splitting
due to $T = 0$ pairing has been observed. This has been
specifically attributed to the attractive matrix elements
of such force. For instance, in the case of $^{118}$Sn,
without $T = 0$ pairing the two QRPA states associated 
with (mainly) the ($\nu {\rm g}_{9/2},\pi {\rm g}_{7/2}$) and
($\nu {\rm h}_{11/2}, \pi {\rm h}_{9/2}$) configurations are split
considerably, the latter configuration being at higher
energy than the former. However, due to the partial
occupation of the $l = 5$ levels which is caused by
the $T = 1$ pairing, they are quite sensitive to $T = 0$
pairing whose attractive matrix elements push the
($\nu {\rm h}_{11/2}, \pi {\rm h}_{9/2}$) downwards: thus, 
this configuration is finally more admixed in the
wave function of the main resonance that it would happen
without $T = 1$ pairing. 

Although these effects are of some interest and could
be detected experimentally in decay experiments, or
other kinds of exclusive experiments, the need is
clear for more direct evidences of $T = 1$ pairing.
This will be the subject of the next subsections.

\subsection{Gamow-Teller states and $T=0$ pair in N$\approx$Z nuclei}

As is discussed above, pairing shows up in the QRPA equations
through its contribution to the p-p matrix elements of the residual
interaction. 
From the viewpoint of the QRPA model, for IAR states with $J^\pi=0^+$, 
only $T=1$ pairing provides a contribution, while for the GT states with $J^\pi=1^+$, 
only $T=0$ pairing contributes; 
for other spin-isospin transitions, such as spin-dipole and spin-quadrupole states,
both $T=0$ and $T=1$ pairing provide contributions.
Therefore, GT states are good candidates to study $T=0$ pairing.

Due to selection rules, the GT transitions connect either single-particle states with 
$j_\nu=j_\pi$, or single-particle 
states with $j_\nu=j_\pi \pm 1$, while the IAR is only made up with the former type of transitions. 
Here, and in what follows, the notation $j_<$ ($j_>$) labels the spin-orbit 
partner with $j_<=l-1/2$ ($j_>=l+1/2$). Usually, the 
former type of transitions $\nu j_>\rightarrow \pi j_>$ or $\nu j_<\rightarrow\pi j_<$ 
is $\approx$ 3-7 MeV lower (higher) in energy than the latter type 
of transitions $\nu j_>\rightarrow\pi j_<$  ($\nu j_<\rightarrow\pi j_>$) due to the spin-orbit
splitting. These considerations would completely govern the unperturbed
nuclear response. RPA correlations, however, play an important role
and we can focus on the matrix elements appearing in the $A$ matrix of
Eq. (\ref{eq:ab_qmatrix}), as they are more important than the matrix
elements that fill up the matrix $B$, as a rule.
In the particle-hole sector the coefficient including occupation
factors reads $u_\alpha v_\beta u_{\alpha^\prime} v_{\beta^\prime}$,
and is non-negligible mainly for configurations 
at relatively high excitation energy.
On the other hand, in the particle-particle sector the 
relevant 
configurations are those made up with partially occupied states near the Fermi surface, that
is, at relatively low excitation energy.
Because of these reasons, the high-lying GT resonance is more sensitive to
the p-h spin-isospin interaction, whose strength has been often related
to the Landau-Migdal parameter $g_0^\prime$~\cite{Bender} as far as the 
central terms are concerned; tensor terms play also a role, in fact~\cite{Bai1,Bai2}. 
For the low energy GT strength, the situation becomes
much more complex~\cite{Engel}: 
in this case, the $T=0$ pairing residual interaction plays a role
and its strength can somehow be pinned down while, however, central and
tensor terms of the p-h residual interaction are also important. 

In order to study the GT transitions in N$\approx$Z nuclei, we have applied
the self-consistent Skyrme HFB+QRPA model. In the calculations, zero-range surface
T=0 pairing was used and its form is the same as it has been shown in Eq. 
(\ref{eq:T1surface}) with a scaling factor $f$, namely
\begin{equation}\label{eq:T0surface}
V_{surface}^{T=0}(\vec r_1,\vec r_2) = \hat{P}_t fV_0 \left( 1 - x\left( \frac{\rho}{\rho_0} 
\right)^{\gamma} \right) \delta\left(\vec r_1 - \vec r_2 \right). 
\end{equation}
where $\hat{P}_t$ is the projector on spin-triplet states.  

\subsubsection{Low energy GT state and $T=0$ pairing in $N=Z$ nuclei \label{LGT}}

\begin{figure}[ht]
\includegraphics[width=\columnwidth]{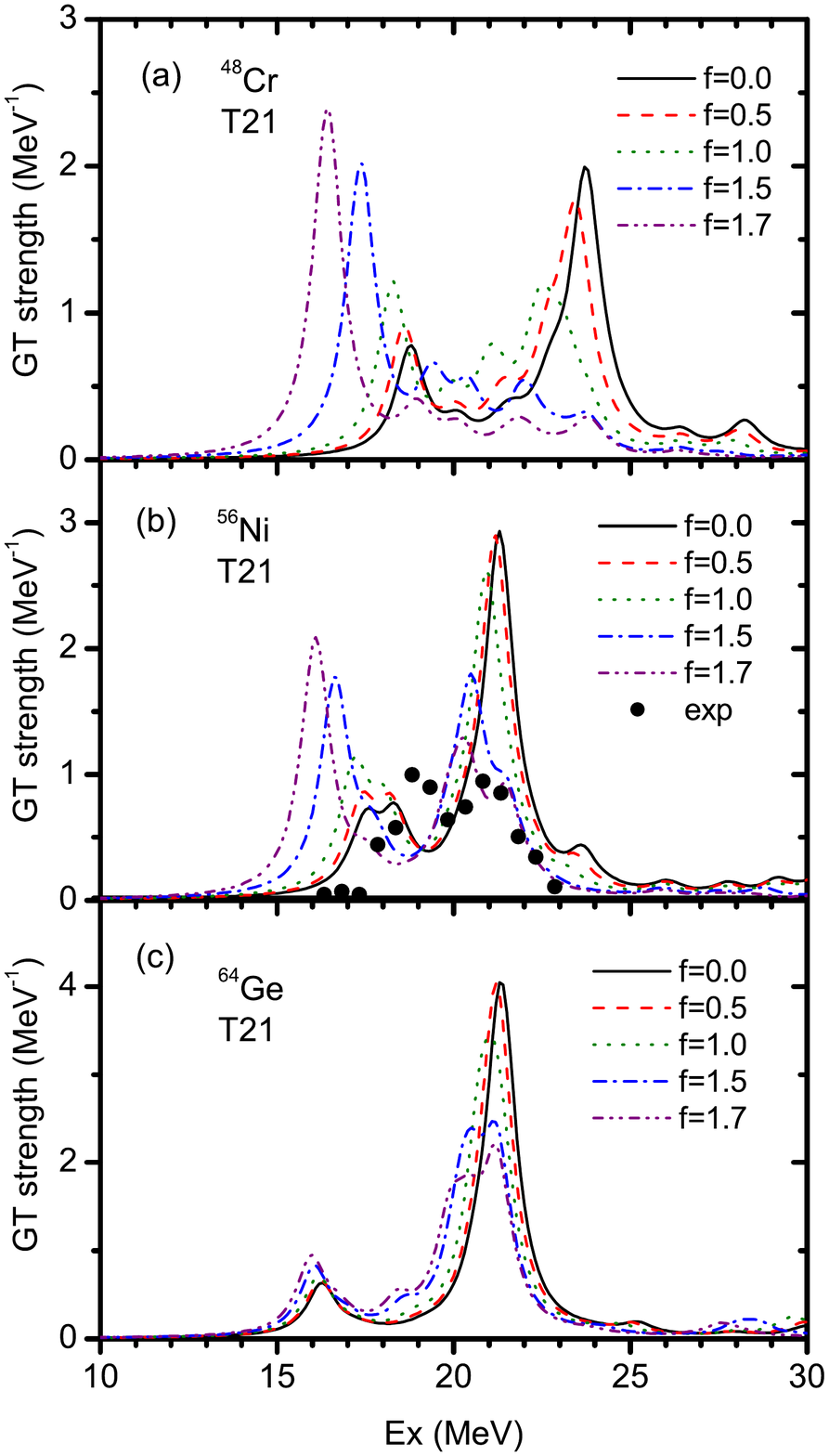}
\caption{\label{fig.2peaks} GT strength distribution in $N=Z$ nuclei $^{48}$Cr, $^{56}$Ni, 
and $^{64}$Ge calculated by HFB+QRPA with the Skyrme force T21 and with different strengths 
of $T=0$ pairing. The excitation energies are calculated with respect to the mother nuclei. 
The experimental data are taken from \cite{Sasano}. The figure is taken from Ref. \cite{Bai13}.}
\end{figure}

We have calculated the GT strength distribution in a series of $N=Z$ $pf$-shell nuclei 
$^{48}$Cr, $^{56}$Ni, and $^{64}$Ge, and the results are shown in Fig. \ref{fig.2peaks}. 
We have performed these QRPA calculations by assuming different strengths
for the $T=0$ pairing or, in other words, the factor $f$ is taken as 0.0, 0.5, 1.0, 1.5, and 1.7 
in Eq. (\ref{eq:T0surface}). The GT strength distribution of 
$^{64}$Ge is shown in the panel (c), and it is evident that with the increasing of the 
$T=0$ pairing strength, a small amount of GT strength is shifted to the low energy region while 
the energy of the low energy peak is shifted downwards albeit only slightly. In 
general only a small amount of GT strength is distributed in this low energy region, and this 
region is about 6 MeV
lower than the high energy peak. This is the normal case of GT strength distributions, as is widely known
in standard nuclei with considerable neutron excess. In the case of $^{56}$Ni, the situation changes: 
when the $T=0$ pairing 
is not included the GT strength distribution is also like in the normal case, i.e. with a main peak 
in the high energy region and
little amount of strength distributed at low energy (actually, if we do HF+RPA calculation, 
without pairing, there is only one peak). 
With the increase of the $T=0$ pairing,
the strength is shifted to the low energy region and meanwhile the low energy peak is shifted downward,
in a more significant fashion than in the previous case. In particular, 
when $f$ = 1.5, there are two peaks having similar amount of strength, and this is qualitatively
consistent with the measured two peak structure~\cite{Sasano}. 
However, as was commented by the authors of Ref. \cite{Machiavelli},
the experimental results are not well reproduced by our present calculation; actually, there 
are other causes 
that may affect the peak energies in $^{56}$Ni, such as the particle-vibration coupling (PVC)
effect \cite{Niu12}. One might
need to use well constrained Skyrme force together with theoretical model that go beyond QRPA, 
like QRPA plus PVC or second QRPA, if one wishes a good reproduction of experimental 
data \cite{GTbench}. For $^{48}$Cr, the situation 
becomes quite extreme, that is, when the $T=0$ pairing strength is strong enough (i.e., with 
$f$ = 1.5 or 1.7), most GT strength is shifted to the low energy region and forms the main peak, 
whereas the original main peak in the high energy region
calculated without $T=0$ pairing disappears. Thus, in this nucleus the $T=0$ pairing might play 
a much more important role.
Therefore, in some $N=Z$ nuclei, the $T=0$ pairing may
play an important role to shift a big amount of GT strength to the low energy region 
to form a strong GT state there. Interestingly, this happens for values of $f$ that
are very consistent with those extracted from shell model matrix elements and discussed
in Ref. \cite{Bertsch1} as well as in our text above.

\subsubsection{Low energy GT state and $T=0$ pairing in $N=Z+2$ nuclei \label{LGT2}}

GT states in a series of $N=Z+2$ $pf$-shell nuclei have been  measured in 
Ref. \cite{Fujita14}, 
and it has been found that while in some nuclei 
the usual high-lying GTR is found, and no significant strength at low energy
appears, in other cases the main GT peak is detected  in the low energy region.
In these cases, the main GT may exhaust more than 80\% of the total strength, and
it has been called Low energy Super Gamow-Teller (LeSGT) state.

\begin{figure}[ht]
\includegraphics[width=\columnwidth]{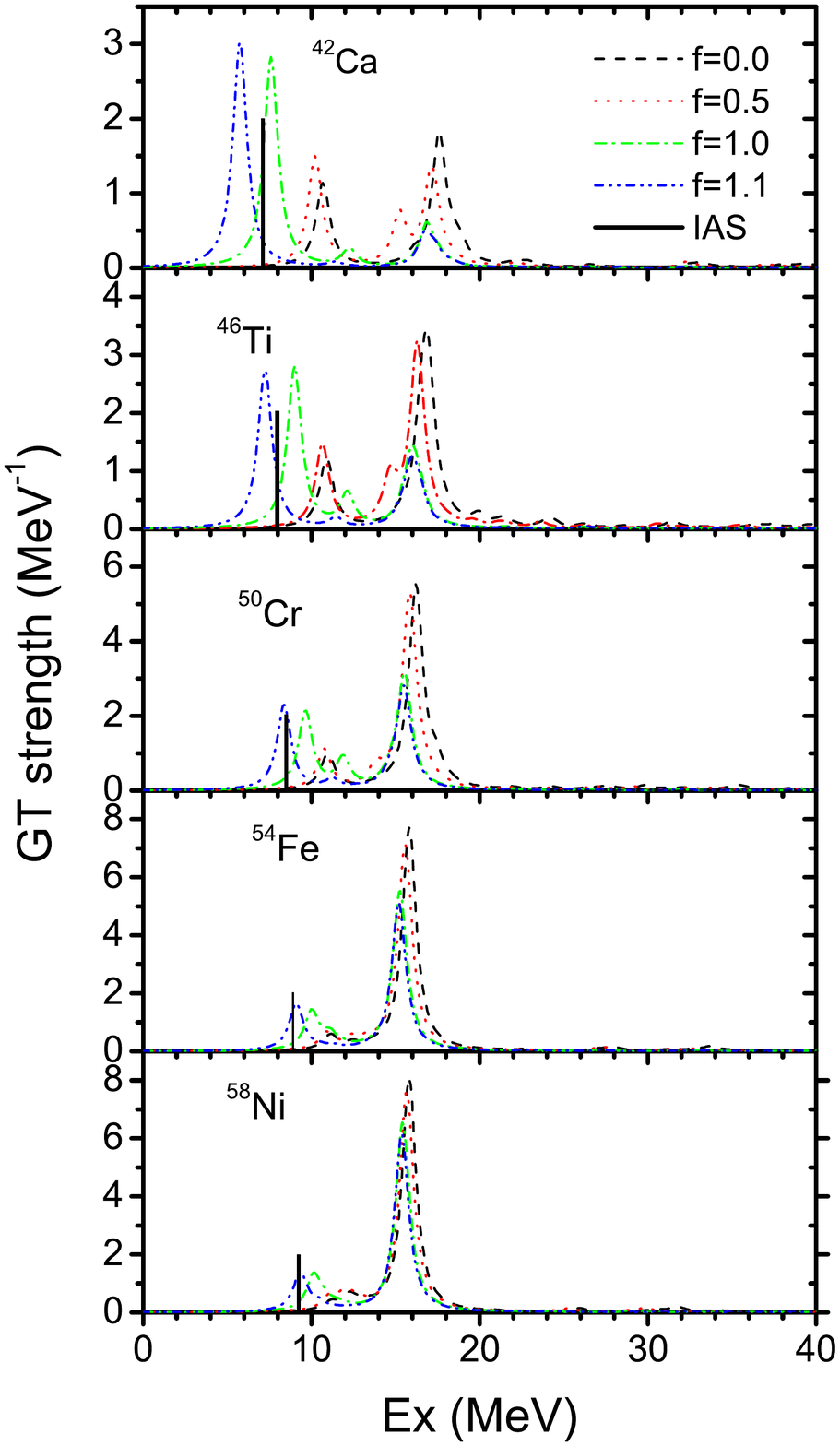}
\caption{\label{fig.SAMi} GT strength distribution in $N=Z+2$ nuclei with mass 
number between 42 and 58, calculated by means of HFB+QRPA with the 
Skyrme force SAMi and different strength of for the $T=0$ pairing. The vertical black line
corresponds to the IAR state. The excitation energies 
are calculated with respect to the mother nuclei. 
The figure is taken from Ref. \cite{Bai14}.}
\end{figure}

The Skyrme HFB+QRPA results for the GT strength distribution in $N=Z+2$ nuclei 
from $A$ = 42 to 58
are shown in Fig. \ref{fig.SAMi}. From this figure, one can see that 
in $^{42}$Ca and $^{46}$Ti the $T=0$
pairing (assumed to have a  strength similar to that of the  corresponding 
$T=1$ pairing) may shift
a big amount of GT strength to the low energy region to form a strong GT state 
(LeSGT).
Due to this reason, the high energy peak disappears. For $^{50}$Cr, $T=0$ pairing 
is also responsible for the shift of 
some amount of GT strength to the low energy region but 
its role is not as strong in the case of $^{42}$Ca and $^{46}$Ti. 
In $^{54}$Fe and $^{58}$Ni, the importance of $T=0$ pairing further decreases.

\begin{figure}[ht]
\includegraphics[width=\columnwidth]{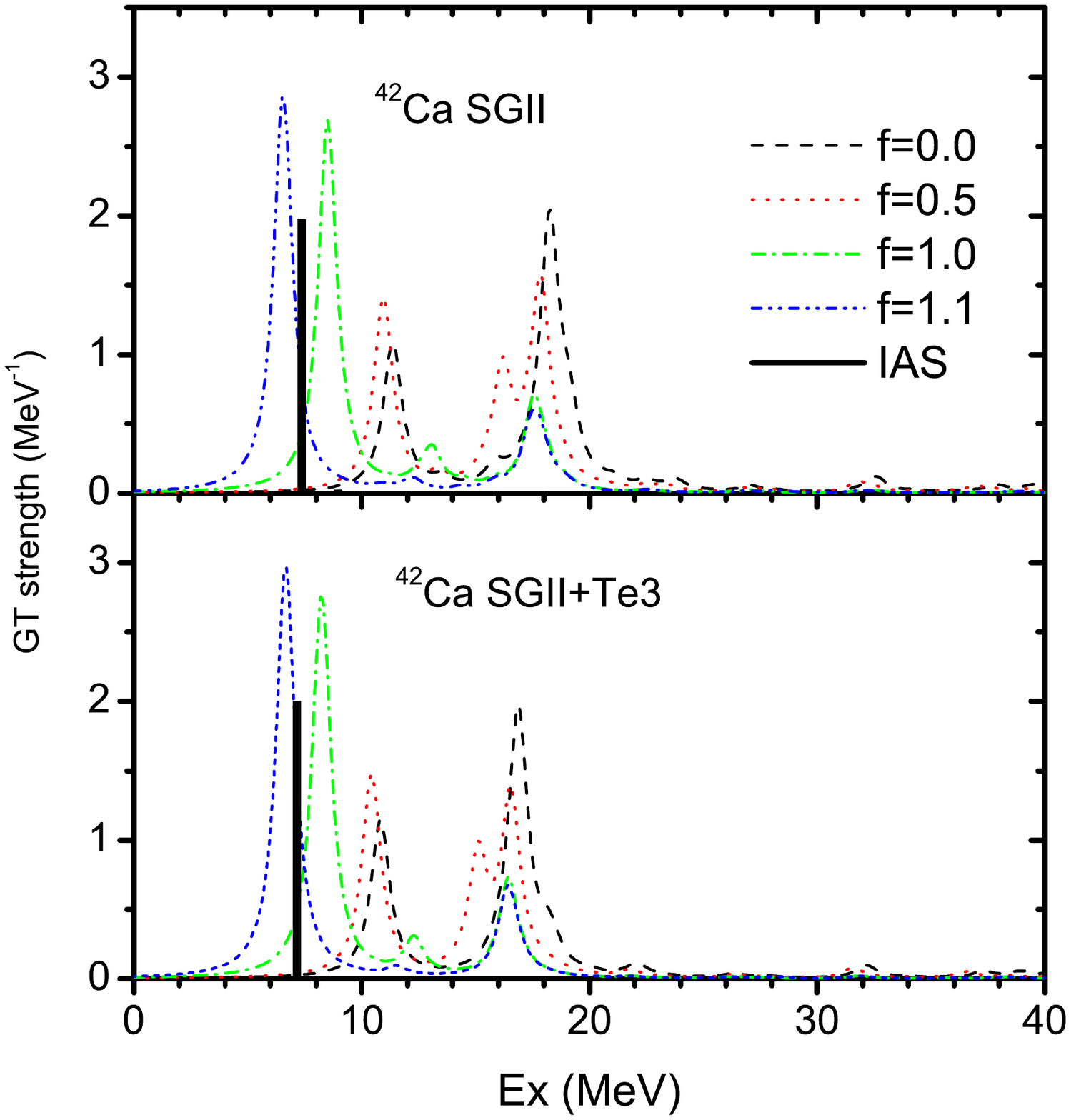}
\caption{\label{fig.Ca} GT strength distribution in $^{42}$Ca calculated by means of 
HFB+QRPA with the Skyrme force SGII, with or without tensor force. The vertical black line
corresponds to the IAR state. The excitation energies
are calculated with respect to the mother nuclei. 
The figure is taken from Ref. \cite{Bai14}.}
\end{figure}

The GT strength distributions in $^{42}$Ca, calculated by means of HFB+QRPA with 
the Skyrme force SGII, with or without the tensor force, are shown in Fig. \ref{fig.Ca}.
In the upper (lower) panel, results without (with) the tensor force, are displayed. 
From this figure one can see that the tensor force plays an important role when 
the $T=0$ pairing
is not strong, i.e. when $f$ = 0.0 and 0.5. Instead, when $f$ takes more realistic 
values like 1.0 or 1.1, the strength 
distributions with and without tensor force are almost the same. This indicates that 
the tensor force effects are suppressed by the $T=0$ pairing with proper strength, and 
this 
means that one can extract reliable information on $T=0$ pairing from such nucleus.

\begin{figure}[ht]
\includegraphics[width=\columnwidth]{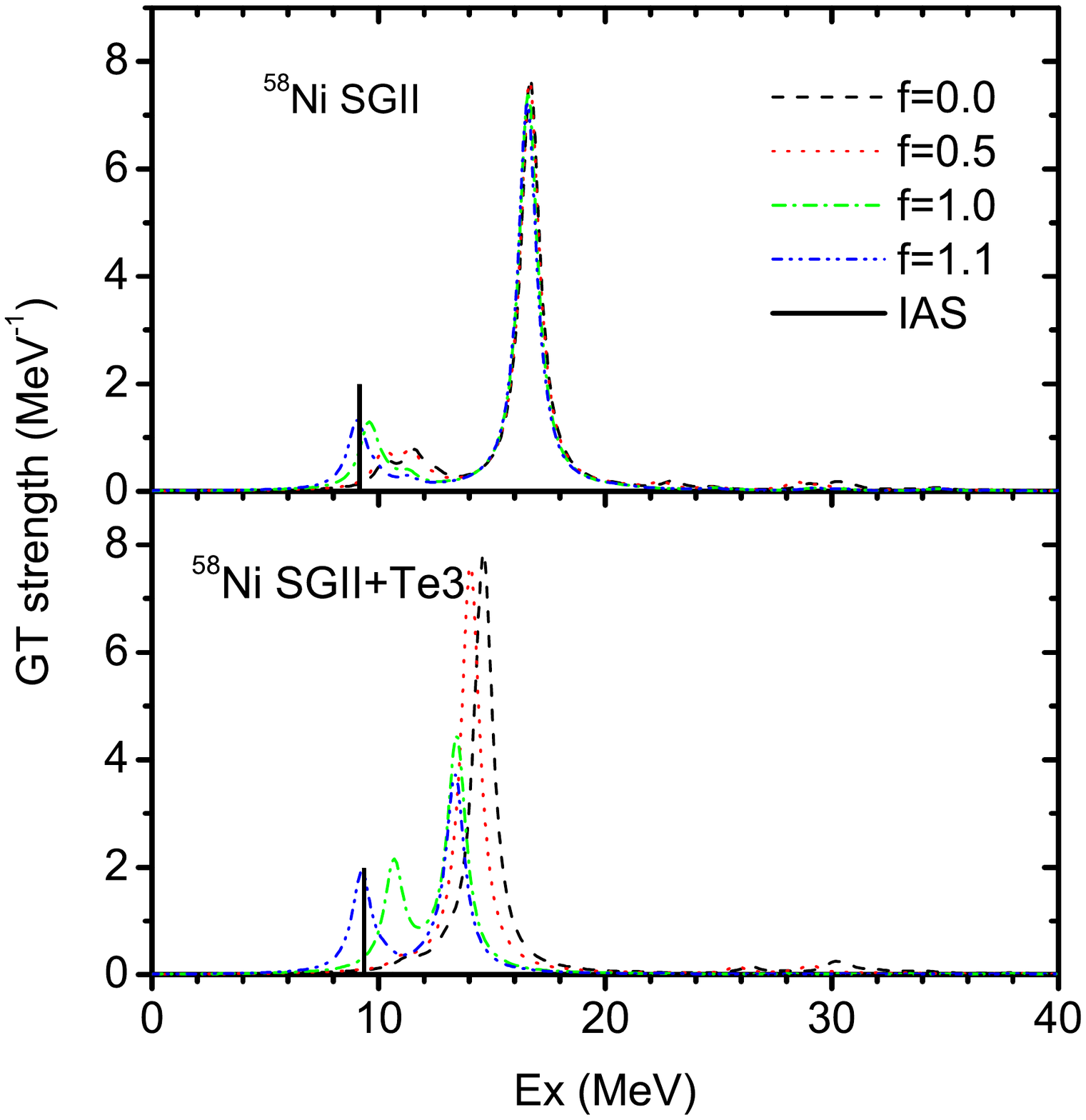}
\caption{\label{fig.Ni} Same as Fig. \ref{fig.Ni} in the case of $^{58}$Ni.
The figure is taken from Ref. \cite{Bai14}.}
\end{figure}

This conclusion may change if another nucleus is chosen.
From Fig. \ref{fig.Ni}, we can see that the GT strength
distributions with and without tensor force are quite different 
in $^{58}$Ni even with
strong $T=0$ pairing strength. This means that in this case we do not have
a unique signature of the effects of the $T=0$ pairing from the low-energy GT 
strength.

Therefore, one can using the low energy GT states in $^{42}$Ca and $^{46}$Ti
to extract reliable information on $T=0$ pairing. Actually, the $T=0$
pairing strength was suggested to be about 1.0 to 1.05 times the strength
of corresponding T=1 pairing in Ref. \cite{Bai14}. 
Although this value is slightly smaller than what we have been discussed in Section \ref{LGT}, 
it has been extracted in a quite direct way. Qualitatively, it confirms that 
$T=0$ pairing is stronger than $T=1$ pairing.

\subsection{Low-lying states in nuclei with one n-p pair outside the core}
\label{Sect.03-2}

We focus in this Section on the energy spectra and spin-isospin excitations of 
odd-odd $N=Z$ $sd$- and $pf$- shell nuclei.
To this end,  
we apply a three-body model with a density dependent 
contact interaction between the valence neutron and proton.  
The three-body model Hamiltonian for odd-odd $N=Z$ nuclei, assuming 
the core + $p + n$ structure \cite{Tanimura12}, 
is given by 
\begin{eqnarray}
H&=&\frac{{\vec{p}_p}^2}{2m}+\frac{{\vec{p}_n}^2}{2m}+V_{pC}
(\vec{r}_p)+V_{nC}(\vec{r}_n) \nonumber \\
&&+V_{pn}(\vec{r}_p,\vec{r}_n)+\frac{(\vec{p}_p+\vec{p}_n)^2}{2A_{C}m},
\label{eq:H}
\end{eqnarray}
where $m$ is the nucleon mass, $A_C$ is the mass number of the core nucleus, 
and $V_{pC}$ and $V_{nC}$ are the mean field potentials 
for the valence proton and 
neutron, respectively, 
generated by the core nucleus. 
These are given as 
\begin{eqnarray}
V_{nC}(\vec{r}_n)=V^{(N)}(r_n),\ V_{pC}(\vec{r}_p)=V^{(N)}(r_p)+V^{(C)}(r_p), 
\label{eq:pot}
\end{eqnarray}
where $V^{(N)}$ and $V^{(C)}$ are the nuclear and the Coulomb parts, 
respectively. In Eq. (\ref{eq:H}), $V_{pn}$ is the pairing 
interaction between the two valence 
nucleons. 
For simplicity, one can safely neglect
the recoil kinetic energy of the core nucleus, that is, 
the last term in Eq. (\ref{eq:H}). 
The nuclear part of the 
core-valence particle interaction, Eq. (\ref{eq:pot}), is taken to be
\begin{equation}
V^{(N)}(r)=v_0f(r)+v_{ls}\frac{1}{r}\frac{df(r)}{dr}(\vec{l}\cdot\vec{s}), 
\label{vnC}
\end{equation}
where $f(r)$ is a Fermi function defined by $ f(r)=1/(1+\exp[(r-R)/a])$. 
For the $^{18}$F nucleus, as in Ref. \cite{Tanimura12}, 
we set $v_0=-49.21$ MeV and $v_{ls}=21.6$ MeV$\cdot$fm$^2$. 
For the 
other nuclei, we adjust $v_0$ so as to reproduce the 
neutron separation energies, while $v_{ls}$ is kept 
constant for all the nuclei. 
For more details on the adopted parameters in the one-body potential, 
the reader can consult Ref. \cite{Tanimura12}.  
We use a contact interaction 
between the valence 
neutron and proton, $V_{np}$, 
that is the sum of $T=0$ and $T=1$ pairing forces with surface character
as defined above in Eqs. (\ref{eq:T1surface}) and (\ref{eq:T0surface}). We write
it here with inclusion of the appropriate projectors for the sake of clarity, namely 
\begin{eqnarray}
V_{np}(\vec{r}_1,\vec{r}_2)&=&\hat{P_s}v_s\delta(\vec{r}_1-\vec{r}_2)
\left(1-x_s\left(\frac{\rho(r)}{\rho_0}\right)^{\gamma}\right)  \nonumber  \\
&+&\hat{P_t}v_t\delta(\vec{r}_1-\vec{r}_2)
\left(1-x_t\left(\frac{\rho(r)}{\rho_0}\right)^{\gamma}\right), \nonumber \\
\label{eq:pairing}
\end{eqnarray}
where $\hat{P}_s$ and $\hat{P}_t$ are the projectors onto the 
spin-singlet and spin-triplet channels, respectively, defined in Eq. (\ref{S-proj}). 
Notice that the isovector spin-singlet pairing 
acts on even-$J$ states $J^\pi=0^+, 2^+, \ldots (2j-1)^+$, 
while the isoscalar spin-triplet pairing acts on
odd-$J$ states  $J^\pi=1^+, 3^+, \ldots (2j)^+$ for the configuration $j^2$.  
In each channel 
in Eq. (\ref{eq:pairing}), 
the first term in brackets
corresponds to the interaction in the vacuum while 
the 
second term takes into account the medium effects through the density 
dependence. Here, the core density is assumed to be 
a Fermi distribution of the same radius and diffuseness as 
in the core-valence particle interaction, Eq. (\ref{vnC}).  
The strength parameters, $v_s$ and $v_t$, are determined 
from the proton-neutron scattering length as \cite{EsBeH97}
\begin{eqnarray}
v_s&=&\frac{2\pi^2\hbar^2}{m}\frac{2a^{(s)}_{pn}}{\pi-2a^{(s)}_{pn}k_{\rm cut}},\\
v_t&=&
\frac{2\pi^2\hbar^2}{m}\frac{2a^{(t)}_{pn}}{\pi-2a^{(t)}_{pn}k_{\rm cut}}, 
\label{eq:v0pair}
\end{eqnarray}
where $a^{(s)}_{pn}=-23.749$ fm and $a^{(t)}_{pn}=5.424$ fm \cite{KoNi75} are 
the empirical p-n scattering lengths in the spin-singlet and spin-triplet 
channels, respectively. 
$k_{\rm cut}$ is the momentum cut-off that must be introduced when treating the 
delta function, 
and is related with the cutoff energy by 
$E_{\rm cut}=\hbar^2k_{cut}^2/m.$ 
In other words, the strengths $v_s$ and $v_t$ determined from 
the scattering lengths depend on the cutoff energy, 
$E_{\rm cut}$. 
The three parameters $x_s,\ x_t$, and $\gamma$ in the density dependent 
terms in 
Eq. (\ref{eq:pairing}) are determined so as to reproduce the 
energies of the ground state ($J^\pi=1^+$), 
the first excited state ($J^\pi=3^+$), and the second excited state 
($J^\pi=0^+$) 
in $^{18}$F 
(see also Ref. \cite{Tanimura12}). 
The Hamiltonian (\ref{eq:H}) is diagonalized in the valence two-particle 
model space. The basis states for this diagonalization are 
given by a product of proton and neutron single 
particle states having single-particle energies $\epsilon^{(\tau)}$:
these energies are obtained with the use of the 
single-particle potential $V_{\tau C}$ in 
Eq. (\ref{eq:H}) ($\tau=p$ or $n$). 
To this end, 
the single-particle continuum states are discretized in a large box.  
We include only those states satisfying 
$\epsilon^{(p)}_{\alpha}+\epsilon^{(n)}_{\beta}\leq E_{\rm cut}$.

\begin{figure}
\begin{center}
\includegraphics[scale=.33,angle=0, bb=0 0 842 595]{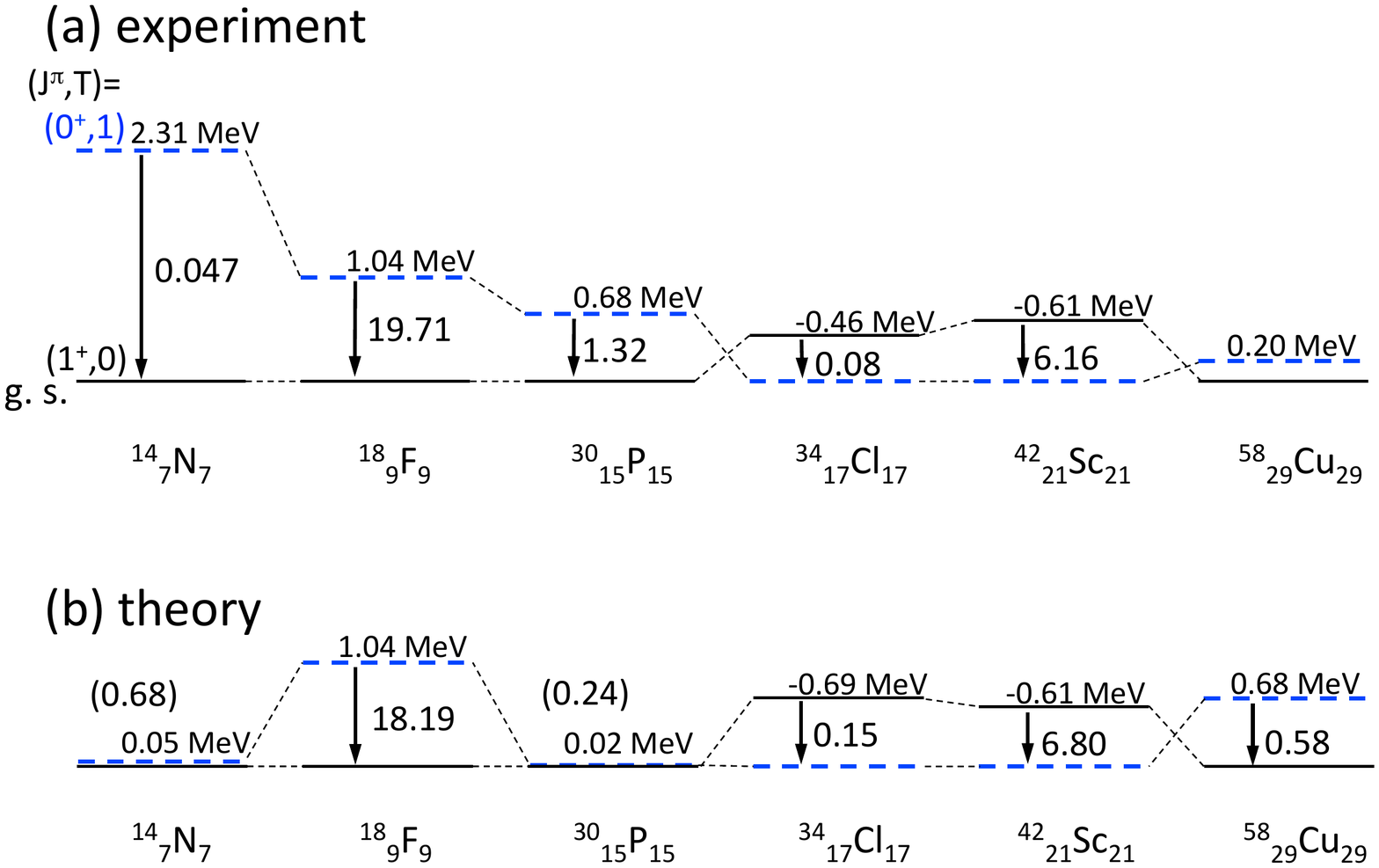}
\end{center}
\caption{(Color online) The energies of the first 0$^+_1$ and the 
first 1$^+_1$ states in $N=Z$ nuclei. The upper panel (a) 
shows experimental data and the lower panel (b) corresponds to 
calculated results.
The values with the arrows 
show the transition probabilities for the  
magnetic dipole transitions, $B(M1)$ (the calculated 
values are shown in brackets for $^{14}$N and $^{30}$P). 
The experimental data are taken from Ref. \cite{exp}.}
\label{fig:levelN=Z}
\end{figure}

The calculated spectra for 
$^{14}$N, $^{18}$F, $^{30}$P, $^{34}$Cl, $^{42}$Sc, and $^{58}$Cu nuclei are 
shown in Fig. \ref{fig:levelN=Z} together with the experimental data.  
The spin-parity for 
the ground state of the nuclei in Fig. \ref{fig:levelN=Z} are 
$J^{\pi}=1^+$ except for 
$^{34}$Cl and $^{42}$Sc. 
This feature is entirely due to the interplay between the isoscalar 
spin-triplet and the isovector spin-singlet 
pairing interactions in these $N=Z$ nuclei.  
In the present calculations, the ratio between the isoscalar and the 
isovector pairing interactions is $v_t/v_s=1.9$ for the energy cutoff 
of the model space $E_{\rm cut}=20$ MeV. 
This ratio is somewhat larger than the value $\approx$ 1.6 extracted in 
Ref. \cite{BL10} from the shell model matrix elements in $p$- 
and $sd$-shell nuclei and already discussed in the previous Sections.
For a larger model space with $E_{\rm cut}=30$ MeV, 
the ratio becomes 1.6, but the agreement 
between the experimental data and the calculations somewhat 
worsens quantitatively even though the general features remain the same.   
It is remarkable that the energy differences 
$\Delta E=E(0^+_1)-E(1^+_1$) are well reproduced in $^{34}$Cl and $^{42}$Sc 
both qualitatively (there is an inversion between the 
0$^+$ state and the 1$^+$ state that becomes the
ground state) and quantitatively (if one looks at 
the absolute value of the energy 
difference). The model description is somewhat poor in $^{14}$N and $^{30}$P
because the cores of these two nuclei are deformed: nevertheless,  
the ordering of the two lowest levels are correctly reproduced. 

The probability of the total spin $S=0$ 
and $S=1$ components 
for the $0^+$ and the 1$^+$ states, respectively, 
are listed in Table \ref{tab:1}.  
The total spin $S=0$ and $S=1$ components in two particle configurations 
can be calculated 
with the formula
\begin{eqnarray}\label{eq:9jtrasf}
 |(j_{\pi}j_{\nu}) J\rangle=\sum_{L,S}\left\{  \begin{array}{ccc}
   l_{\pi} & l_{\nu}    &L  \\
   s  &  s &   S  \\
   j_{\pi} &  j_{\nu} & J   \end{array} \right \}     \hat{L} \hat{S} \hat{ j_{\pi} }\hat{ j_{\nu} } 
   |(  l_{\pi}  l_{\nu} )LS;J\rangle,   \nonumber \\
   \end{eqnarray}
where $\hat{L}, \hat{S}, \hat{j}$ are $\hat{L}=\sqrt{2L+1}, \hat{S}=\sqrt{2S+1}, \hat{j}=\sqrt{2j+1}$, 
respectively.
For a $ j_{\pi} = j_{\nu} =j=l+1/2$ configuration, 
the $S=0$ and $S=1$ components are given by the factors 
$(j+1/2)/2j$ and  $(j-1/2)/2j$, respectively, 
for $J=0$. 
For a $ j_{\pi} = j_{\nu} =j=l-1/2$ configuration, 
on the other hand, they are 
$(j+1/2)/(2j+2)$ and $(j+3/2)/(2j+2)$ for $S=0$ and $S=1$, respectively. 
Notice that the $s_{1/2}^2$ configuration has only $S=0$ component if $J=0$. 
Otherwise, all the two particle states have a 
large mixture of the $S=0$ and $S=1$ components.
In general, the $S=1$ and $S=0$ components are thus 
largely mixed in the wave 
functions of both the ground and the excited states. 
An exception is $^{30}$P. 
In this nucleus, the dominant configuration in the 0$^+$ state 
is $(2s_{1/2}^{\pi}\otimes 2s_{1/2}^{\nu})$, 
which can couple only to the total spin $S=0$. 
On the other hand, in the 1$^+$  state, the dominant 
configuration is 
$(2s_{1/2}\otimes 1d_{3/2})~{T=0}$ which can couple only to 
the total spin $S=1$ 
with the total angular momentum $L=2$.

\begin{table*}
\caption{The energy difference between the 0$^+_1$ and $1^+_1$ states, 
$\Delta E=E(0^+_1)-E(1^+_1)$, in $N=Z$ nuclei. The probabilities of the $S=0$ 
component $P(S=0)$ in the wave functions for the $0_1^+$ state 
are shown in the fourth line. The fifth line shows the probability of the 
$S=1$ 
component in the 1$^+$ state. 
The probabilities $P(j^{\pi}\otimes j^{\nu})$ for the 
dominant valence shell proton-neutron configuration are 
also given for the 0$^+_1$ and 1$^+_1$ states in the seventh and 
eigth line, 
respectively.  
The experimental data are taken from Ref. \cite{exp}.}
\begin{center}
\begin{tabular}{cc|cccccc}
\hline\hline
 & & $^{14}$N & $^{18}$F & $^{30}$P & $^{34}$Cl & $^{42}$Sc & $^{58}$Cu \\
\hline
$\Delta E$ & exp. & 2.31 & 1.04 & 0.68 & $-0.46$ & $-0.61$ & 0.20 \\
(MeV) &                 cal. & 0.05 & 1.04 & 0.02 & $-0.69$ & $-0.61$ & 0.68 \\
\hline
$P(S=0)$ (\%) & $0^+$           & 34.8 & 82.2& 94.8 & 40.7 & 70.5 & 65.4 \\
$P(S=1)$ (\%) & $1^+$           & 78.3 & 90.1 & 95.8 & 64.3 & 65.7 & 92.1 \\
  \hline
\multicolumn{2}{c|}{$j$}               & $1{\rm p}_{1/2}$ & $1{\rm d}_{5/2}$ 
& $2{\rm s}_{1/2}$ & $1{\rm d}_{3/2}$ & $1{\rm f}_{7/2}$ & 
$2{\rm p}_{3/2}$ \\\hline
$P(j^{\pi}\otimes j^{\nu})$   & $0^+$          
& 97.2 & 85.2 & 89.7 & 98.6 & 94.2 & 81.2 \\
(\%) & $1^+$      & 96.4 & 52.1 & 1.1 & 98.4 & 82.7 & 10.0 \\
\hline\hline  \label{tab:1}
\end{tabular}
\end{center}
\end{table*}

The reduced magnetic dipole transition probability is given by
\begin{eqnarray}
&&B(M1:J_i \rightarrow J_f) =  \nonumber \\
&& \left(\frac{3}{4\pi}\right)\frac{1}{2J_i+1}\left|\langle 
J_f||\sum_{i}(g_s(i)\vec{s}_i+g_l(i)\vec{l}_i)||J_i\rangle\right|^2,
\nonumber 
\\
\label{eq:m1}
\end{eqnarray}
where the double bar means a reduced matrix element in the spin space.    
We take the bare $g$ factors $g_s(\pi)=5.58 $, $g_s(\nu)=-3.82$, 
$g_l(\pi)=1$, and $g_l(\nu)=0$ for the magnetic moments, and the magnetic 
dipole transitions are given in 
unit of the nuclear magneton $\mu_N=e\hbar/2mc$. 
The spin-quadrupole transition is defined instead by
\begin{eqnarray}
&&B(IVSQ:J_i \rightarrow J_f) =  \nonumber \\
&& \frac{1}{2J_i+1}\left|\langle J_f||\sum_{i} r_i^2[\vec\sigma(i)Y_2(i)]
^{(\lambda=1)} \tau_z(i) ||J_i\rangle\right|^2. \nonumber \\
\label{eq:IVSQ}
\end{eqnarray}

The calculated magnetic moments and magnetic dipole transitions 
are listed in Table \ref{tab:2} together with the spin quadrupole transitions.
The calculated magnetic moment in $^{14}$N reproduces well the observed one, 
while the agreement is 
worse 
in $^{58}$Cu.
This is due to the fact that the core of $^{56}$Ni might be largely broken 
and the $f_{7/2}$ hole configuration is mixed in the ground state of $^{58}$Cu 
~\cite{Lisetskiy2003,Honma2004}.
The values for $B(M1)$ are also shown in Fig. \ref{fig:levelN=Z}. 
Very strong $B(M1)$ values are found 
both experimentally and theoretically in two of the $N=Z$ nuclei 
in Table \ref{tab:2}, that is, in $^{18}$F and $^{42}$Sc. 

The $B(M1)$ transition 
from 0$^+$ to 1$^+$ in 
$^{18}$F is the largest one so far observed in the entire 
region of nuclear chart. 
We notice that our three-body calculations provide fine agreements 
not only for these strong transitions in $^{18}$F and $^{42}$Sc but 
also for the quenched transitions in the other $N=Z$ nuclei such as in 
$^{14}$N and $^{34}$Cl.  
The shell model calculation of Ref. \cite{Brown-W}
shows also a large enhancement for the $B(M1)$ transition 
in $^{18}$F which is consistent with both the present study and 
the experiment.

\begin{table*}
\caption{The 
magnetic dipole transitions, and the 
isovector spin quadrupole transitions in the $N=Z$ nuclei are displayed.
The experimental data for the $B(M1)$ values are taken from Ref. \cite{exp}.
The symbol $\downarrow (\uparrow)$ denotes
a transition from the excited (ground) state to the ground (excited) state.}
\begin{center}
\begin{tabular}{cc|cccccc}
\hline\hline
 & & $^{14}$N & $^{18}$F & $^{30}$P & $^{34}$Cl & $^{42}$Sc & $^{58}$Cu \\
\hline
$J^{\pi}_{
gs}$ & & $1^+$ & $1^+$ & $1^+$ & $0^+$ & $0^+$ & $1^+$\\
\hline
 $B({
 M}1)\downarrow
 $   ($\mu_N^2$)   & exp. & 0.047 & 19.71 & 1.32 & 0.08 & 6.16 & -- \\
                                & cal. & 0.682 & 18.19 & 0.24 & 0.15 & 6.81 & 0.580 \\
\hline
 $B({
 SQ})\uparrow
 $ (fm$^4$) & cal. & 33.17 & 0.85 & 43.04 & 74.52 & 19.61 & 71.55 \\
 \hline\hline  \label{tab:2}
\end{tabular}
\end{center}
\end{table*}
  
In the case of $^{18}$F, the $0^+$ and $1^+$ states are 
largely dominated by the $S=0$ and $S=1$ 
spin components, respectively, with orbital angular momentum $l=2$ (see 
Table \ref{tab:1}). 
Therefore, the two states can be considered as members of 
the SU(4) multiplet 
in the spin-isospin space \cite{Wigner37}. 
This is the main reason why the $B(M1)$ value
is so large in this nucleus, 
since the spin-isospin operator $g_s^{IV}\vec{s}\tau_z$ 
connects two states in the same SU(4) multiplet, 
that is, the transition is allowed, 
and the isovector $g$-factor is the dominant term in Eq. (\ref{eq:m1})
with $g_s^{IV}=(g_s(\nu)-g_s(\pi))/2=-4.70$. 
The configurations in $^{42}$Sc are also similar to 
those in $^{18}$F in terms of SU(4) multiplets, 
although they are dominated by $l=3$ wave functions.  
For $^{14}$N and $^{34}$Cl, 
the $B(M1)$ transitions do not acquire any enhancement, since 
the $S=0$ component in the $0^+$ state 
is suppressed due to the $j=l-1/2$ coupling:
both the $0^+$ and $1^+$ states have very large 
$1p_{1/2}^2$ ($1d_{3/2}^2$) configurations in $^{14}$N ($^{34}$C).
These indications for the SU(4) symmetry in $^{18}$F and $^{42}$Sc are 
consistent with the results obtained in Refs. \cite{HaBa89,VoOr93,IsWaBr95}. 

In the nuclei $^{30}$P and $^{58}$Cu, the 1$^+$ state is 
dominated by 1$d_{3/2}2s_{1/2}$ and 
$2p_{3/2}1f_{5/2}$ configurations, respectively, while the 0$^+$ 
state is governed by 
the $2s_{1/2}^2$ and $2p_{3/2}^2$ configurations, respectively. 
Therefore the isovector spin-quadrupole transitions 
are largely enhanced in the two nuclei even though 
the $B(M1)$ is much quenched.  
The validity of SU(4) symmetry has been known already 
for a quite long time 
in $p-$shell shell nuclei \cite{SU4}.
The two-body matrix element of Cohen-Kurath \cite{Cohen-Kurath} 
for the spin-triplet  $(J,T)=(1,0)$ interaction 
is certainly stronger than that for   
the spin-singlet $(J,T)=(0,1)$ pairing interaction.  
Then, the structure of the two-body wave function will be rather 
described by the $LS$ 
coupling scheme than ${jj}$ coupling scheme.

\subsection{Gamow-Teller transitions in $N\approx Z$ nuclei by the 
three-body model}
\label{Sect.03-3}

Some of us have applied the same three-body model as 
described in Subsection \ref{Sect.03-2} to calculate the GT 
strength of $N=Z+2$ nuclei.
The Gamow-Teller (GT) transition strength reads 
\begin{eqnarray}
B(GT:0^+ \rightarrow 1^+) = 
\frac{g_A^2}{4\pi}\left|\langle1^+||\sum_{i}t_-(i)\vec\sigma(i)||0^+
\rangle\right|^2,
\label{eq:GT}
\end{eqnarray}
where $g_A$ is the axial-vector coupling constant. The results are  
summarized 
in Table \ref{tab:GT}. 
One can again see a strong GT transition between the 
lowest $0^+$ and $1^+$ states in $A$ = 18 and 42 systems, 
which exhausts a large portion of the GT sum rule value.  
This can also be interpreted as 
a manifestation of SU(4) symmetry in the wave functions of these nuclei. 
We note also that the result obtained in Ref. \cite{HaBa89} by an analysis of 
GT transitions also implies a good SU(4) symmetry in the $A$ = 18 system. 
On the other hand, for $^{58}$Cu, 
the GT strength is largely fragmented and 
no state with a strong $B(GT)$ is seen near the ground state. 
The experimental data are 
consistent with the calculated results as can be seen in Table \ref{tab:GT}. 
The ratio of $B(GT)$ values from the ground states, between $A$ = 18, 42 
and 58, is 1:0.69:0.05 if one takes the experimental values while the 
calculated ratio is 1:0.71:0.04 and is is very close to the experimental one.  
This agreement suggests the validity of the three-body model 
wave functions in these nuclei.

\begin{table}
\caption{The Gamow-Teller transition probabilities from the ground states 
of $^{18}$O to $^{18}$F, $^{42}$Ca to $^{42}$Sc, and $^{58}$Ni 
to $^{58}$Cu, 
in units of $g_A^2/4\pi$. 
The experimental data are taken from Ref. \cite{GT1} for $^{18}$F, Ref. 
\cite{GT2} for $^{42}$Sc and Ref. \cite{GT3} for $^{58}$Cu, respectively.}
\begin{center}
\begin{tabular}{cc|cc}
\hline\hline
 \multicolumn{4}{c}{$^{18}$O $\rightarrow ^{18}$F} \\
\hline 
 \multicolumn{2}{c|}{$E_x$ (MeV)} & 
 \multicolumn{2}{c}{$B(GT)$ ($g_A^2/4\pi$)} \\
\hline
 Th. & Exp.    &    Th. & Exp.  \\
\hline
 0.0 & 0.0      &      2.48 & 3.11$\pm$ 0.03  \\
 4.79 &     &     0.028 & \\
 6.87 &  &       0.036  & \\
\hline
 \multicolumn{4} {c}{$^{42}$Ca $\rightarrow ^{42}$Sc}   \\\hline 
 \multicolumn{2}{c|}{$E_x$ (MeV)} & \multicolumn{2}{c}{$B(GT)$ ($g_A^2/4\pi$)} \\
\hline
 Th. & Exp.    &    Th. & Exp.  \\
\hline 
  0.61 & 0.61   &   1.80 & 2.16  $\pm$ 0.15  \\
   & 1.89   &     &0.09 $\pm$ 0.02  \\
  3.71 & 3.69  &   0.346 & 0.15 $\pm$ 0.03 \\
\hline
 \multicolumn{4} {c}{$^{58}$Ni $\rightarrow ^{58}$Cu}   \\
\hline 
 \multicolumn{2}{c|}{$E_x$ (MeV)} & \multicolumn{2}{c}{$B(GT)$ ($g_A^2/4\pi$)} \\
\hline
 Th. & Exp.    &    Th. & Exp.  \\
\hline   
 0.0 & 0.0   &     0.097 & 0.155  $\pm$ 0.01 \\
 1.24 &1.05  &    0.74 & 0.30  $\pm$   0.04  \\
\hline\hline
\label{tab:GT}
\end{tabular}
\end{center}
\end{table}

Extensive shell model calculations have been performed in the full 
$p$-shell, $sd$-shell and $pf$-shell 
model spaces in the literature 
(cf. Refs. \cite{Cohen-Kurath}, \cite{Brown-W} and \cite{Honma2004},  
respectively).
In these studies, the magnetic moments, M1 transitions and GT transitions 
were studied 
and the calculated results reproduce well the experimental 
observations. The validity of SU(4) symmetry in GT decays 
was also studied in terms of shell model calculations 
of $p$-shell nuclei in Ref. \cite{Millener}.  
In contrast, 
our aim in this paper is not to compete with these complete 
shell model calculations, but to extract the role of the spin-singlet 
and the spin-triplet pairing interactions for the ground states 
and the excited states in the odd-odd $N = Z$ nuclei by the three-body 
model with one set of input data for the entire mass region, and 
to explore therefore the validity of SU(4) 
symmetry in the spin-isospin space simple terms.
We should note 
that the present model is quite appropriate for $^{18}$F 
and $^{42}$Sc since $^{16}$O and $^{40}$Ca are doube closed-shell  
nuclei and can be considered as good cores. 
On the other hand, the model space of the three-body model 
is not quite large enough for $^{30}$P, $^{34}$Cl and 
$^{58}$Cu 
since excited states of the core might be coupled 
to the configurations of the present model space. 

For $^{14}$N, the two-hole three-body model might be more appropriate 
to adopt since $^{16}$O is a better core than $^{12}$C. With the 
density dependent, surface-type pairing [$x_s =x_t =+1$ in Eq. (\ref{eq:pairing})], 
the energy difference between the 0$^+$ and 1$^+$ states 
becomes $\approx$ 2 MeV which is close to the experimental observation. 
However, the magnetic dipole transition becomes much larger than the 
observed one. It can be pointed out that the mixing of $sd$ shell 
components will play an important role for the 
quenching of B(M1) and B(GT) values.

\subsection{Pair transfer reactions}
\label{Sect.03-5}

As it is intuitive and well known, two-particle transfer
reactions are sensitive to the correlations between those
particles, so that two-neutron transfer has been for
long time used to pin down fingerprints of $T = 1$ pairing,
and recently there is a strong interest to understand
to which extent deuteron transfer can probe $T = 0$ pairing.

There are some basic problems, nonetheless, both at the conceptual
and experimental level. Let us assume we can restrict to $L = 0$ states
excited in transitions between even-even nuclei. Normal QRPA, or
shell-model calculations, can provide the wave functions of such excited
states $n$; in particular, one can calculate the strength
functions $S(E)$ associated either with the pair addition (ad) or pair removal (rm)
operators. These strength functions read
\begin{eqnarray}\label{eq:pairS}
S_{\rm ad} & = & \sum_{n \in {A+2}} \vert \langle n \vert {\hat P}^\dagger \vert 0 
\rangle \vert^2\ \delta(E-E_n), \nonumber \\
S_{\rm rm} & = & \sum_{n \in {A-2}} \vert \langle n \vert {\hat P} \vert 0 \rangle 
\vert^2\ \delta(E-E_n),
\end{eqnarray}
where the sums run over the appropriate set of final states in the $A+2$ or
$A-2$ nuclei, and $\vert 0 \rangle$ is the ground state of the even-even
nucleus under study. The pair addition operators is
\bea
&&{\hat P}^\dagger_{T=1,S=0\ (T=0,S=1)} = \nonumber \\
&&\frac{1}{2}\sum_{\sigma,\sigma'}\sum_{\tau,\tau'}
\int d^3r\ {\hat \psi}^{\dagger}({\vec r} \sigma \tau) \langle
\sigma \tau \vert \tau_z (\sigma_z) \vert \sigma' \tau' \rangle
{\hat \psi}^{\dagger}({\vec r} {\bar\sigma}' {\bar\tau}'), \nonumber \\
\label{eq:np-pair}
\eea
where ${\hat \psi}^{\dagger}({\vec r} \sigma \tau)$ is the nucleon field operator,
and the time-conjugate state is defined by
${\hat \psi}^{\dagger}({\vec r} {\bar \sigma} {\bar \tau}) \equiv (-2\sigma)(-2 \tau)
{\hat \psi}^{\dagger}({\vec r} -\sigma -\tau)$. We have specifically singled
out the $S,T$ dependence. One can consider the neutron-neutron case 
\cite{Matsuo3}
where only $J^\pi = 0^+, T=1, S=0$ is possible if $L=0$,
or the neutron-proton
case \cite{Yoshida} where either $J^\pi = 0^+, T=1, S=0$ or $J^\pi = 1^+, T=0, S=1$ are
possible.

The strength functions (\ref{eq:pairS}) are known to be 
characterized by collective modes like pair vibrations and
rotations \cite{Brink1}. A so-called giant pairing vibration
(GPV) has been predicted long ago \cite{Bes}, and yet it has not
been so far unambiguously identified despite many efforts that have 
been carried out since the 1970s until very recently (cf. \cite{Mouginot} 
and references therein). One of the reasons that have been advocated 
is that GPVs are expected to lie at high energies, and the semi-classical 
condition of the optimal $Q$-value favors ground-state to ground-state 
transitions in the reactions with stable beams and targets (cf. \cite{Fortunato} 
and references therein). 

The interest has recently switched 
to the behaviour of such excitations far from stability,
and to the issue whether this can bring information on
pairing in neutron-rich nuclei. The strength 
functions (\ref{eq:pairS}) have been calculated
for the Sn isotopes in Ref. \cite{Matsuo3}. It has been
found that ground-state to ground-state transitions have
large pair addition strength in $^{110-130}$Sn. In
$^{132-140}$Sn, characteristic pair vibrational modes
show up, whose strength is slightly smaller than that
of the ground-state transitions but still large enough in terms
of matrix elements associated with single-particle pair transfer.
In these nuclei which are considerably neutron-rich,
such vibrational modes have an extended radial tail, associated
with contributions from continuum transitions. A signature of
this fact resides in the participation of high angular
momentum states with $l \ge 5$ in these modes. Pairing
rotational modes show up instead in more neutron-rich nuclei.
These states, and especially their collectivity, is very sensitive
to the kind of pairing interaction that is employed like
volume pairing as in (\ref{eq:vol_pair}) or surface pairing 
like in (\ref{eq:T1surface}).

However, the pairing addition or pairing removal strength is
not directly observable. The measurable quantity, namely
the two-particle transfer cross section is not simply a factorized
product of such matrix element times kinematical factors since
the reaction process is quite involved. 
The full microscopic theory of the reaction process is explained 
in textbooks \cite{Glendenning} and review papers \cite{Vitturi}. 
Sophisticated calculations of absolute cross sections
have been published starting from the 1960s (see e.g. 
\cite{Thompson} for a compact r\'esum\'e and a survey of references), until the very recent
state-of-the-art scheme of Ref. \cite{Potel}.
We do not wish here to give a full account of these calculations but simply
convey the main points that are also connected to the similar
difficulties that one may expect in the case of  
$T = 0$ pairing.

\begin{figure}
\vspace{-3cm}
\includegraphics[scale=0.35,angle=0, bb=0 0 720 540]{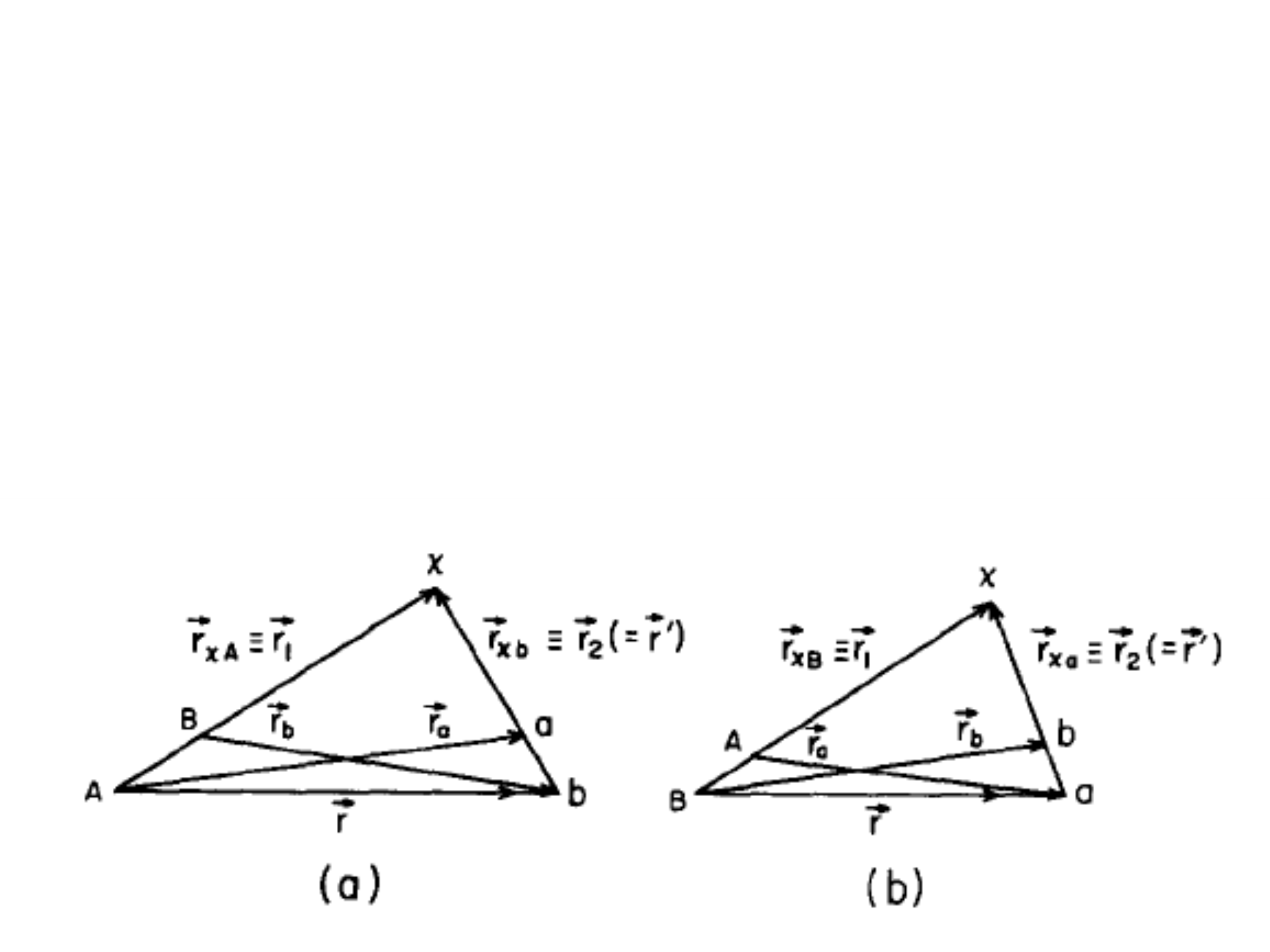}
\caption{Choice of coordinates appropriate for (a) stripping and (b)
pick-up reactions. Figure taken from Ref. \cite{Tamura}.}
\label{fig:tr_coord}
\end{figure}

In a 
Distorted Wave Born Approximation (DWBA) picture, the reaction
cross section associated with $A+a\rightarrow B+b$ will be proportional to the square of the 
transition amplitude. At variance with the inelastic case, the
nucleons are in different partitions in the initial $Aa$ and final
$Bb$ channels. One could define a {\em natural} coordinate system
in each of these two initial and final channels, and in this way 
identify the two relative coordinates ${\vec r}_a$ and ${\vec r}_b$
in such channels (see e.g. Refs. \cite{Satchler,Tamura} for details
and Fig. \ref{fig:tr_coord} for illustration). The 
projectile-target interaction can be defined either in the
initial or final channel (where the words ejectile-residual could be
more appropriate), and one refers to the two choices as prior or
post representations, respectively. In case of a finite-range
interaction $V$ such features brings in many complications associated 
with non-localities. Neglecting them,
that is, approximating ${\vec r}_a \approx {\vec r}_b \approx {\vec R}$ 
as in Eq. (17) of \cite{Vitturi}, the transition amplitude reads
\begin{equation}
T_{A+a\rightarrow B+b} \approx 
\int d^3R\ \chi^\dagger_{Bb}({\vec r}_a, {\vec k}_f) F(\vec R) 
\chi_{Aa}({\vec r}_b, {\vec k}_i),
\label{TA}
\end{equation}
where the $\chi$ are distorted wave functions that carry 
appropriate momentum labels, while $F$ is the reaction form
factor,
\begin{equation}\label{eq:formfactor}
F(\vec R) \approx \langle \Psi_B \Psi_b \vert V \vert \Psi_A \Psi_a \rangle,
\end{equation}
which is written in terms of intrinsic wave functions $\Psi$. We remind
that the differential cross section is 
\begin{equation}
\left( \frac{d\sigma}{d\Omega} \right) = \vert T_{A+a\rightarrow B+b} \vert^2. 
\end{equation}
What appears clearly, then, is that 
while 
the matrix elements that appear in Eq. (\ref{TA}) 
 involve 
a full integration over 
space, 
in the form factor (\ref{eq:formfactor}) the effective interaction $V$ 
is active only in the range allowed by the reaction mechanism and 
acts a kind of filter that makes the connection between pairing correlations
and reaction cross sections quite indirect. 

In such a complicated situation, the comparison of the latest theoretical 
calculations of \cite{Potel} with the experimental data from Ref. 
\cite{Guazzoni} have nonetheless provided strong evidence that our current 
understanding of $T = 1$ pairing is confirmed by the analysis of the 
results of (p,t) transfer reactions on the stable Sn isotopes.
On the other hand, reaction cross section calculations have not
been performed for neutron-rich and weakly bound isotopes e.g.
beyond $^{130}$Sn. At present, such very neutron-rich nuclei are not yet
available in such intensities that allow two-particle transfer reactions
experiments but the topic is certainly of interest for the future.

It is quite obvious also to analyze whether neutron-proton 
transfer reactions can play a similar role as discussed so far, to pin
down better evidences for $T = 0$ pairing. The issue has been recently
discussed in Ref. \cite{Machiavelli}. One clear physics case would be 
to perform a reaction like ($^3$He, p) on an even-even $N = Z$ nucleus. 
$N = Z$ nuclei are stable only up to $^{40}$Ca. As the pairing collectivity is
expected to be more pronounced for heavier systems, experimental programs 
for ($^3$He, p) or ($^4$He, d) 
reactions with unstable beams in inverse kinematics are strongly desired 
and called for. In such nuclei
the $T = 0$ pairing correlations are expected to be enhanced as already
discussed. Moreover, in that case one starts from a $T = 0$, $J^\pi = 0^+$ state
and can probe the isospin invariance of $T = 1$ pairing by looking at
the $T = 1$, $J^\pi = 0^+$ states in the odd-odd system and investigate
$T = 0$ pairing by looking at the $T = 0$, $J^\pi = 1^+$ states. 

A full theory like the one developed for equal-particle transfer reactions, that
is, capable to predict absolute values of the cross section as we have
just discussed, need still to be developed. This is one of the important
priorities in the nuclear reaction domain. Meanwhile, if experimentally
available, relative cross sections $\sigma(T=0,J=1^+)/\sigma(T=1,J=0^+)$
can provide first valuable information. It has to be noted that in such
cross sections, strong interaction matrix elements that are exactly the
analogous of the electromagnetic ones discussed in Sec. \ref{Sect.03-2},
play the important role.

\begin{figure}
\includegraphics[scale=.35,angle=0, bb=0 0 720 540]{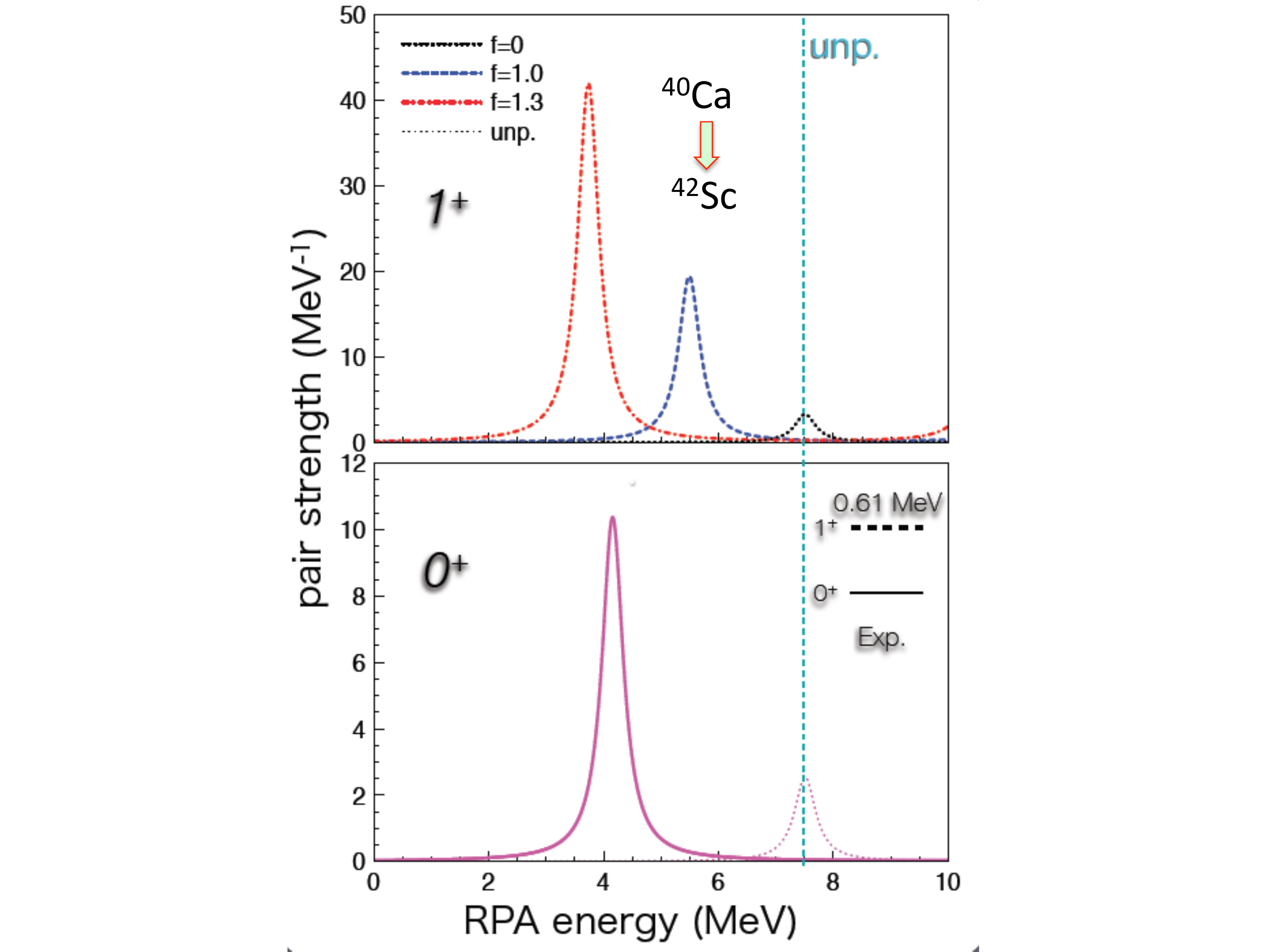}
\caption{(Color online) $L=0$ neutron-proton pair addition 
strength (\ref{eq:pairS}) in the case of the transitions 
$^{40}$Ca $\rightarrow$ $^{42}$Sc. In the case of $J^{\pi}=1^+$ 
states the operator ${\hat P}_{T=0,S=1}$ is active, whereas in the 
case of $J^{\pi}=0^+$ states the strength is associated with 
${\hat P}_{T=1,S=0}$. The sharp peaks associated with the strength 
function (\ref{eq:pairS}) are smeared by means of Lorenzian 
functions with a width of 0.1 MeV. In the $(J,T)=(1,0)$ channel the 
spin-triplet pairing strength is changed by scaling factors 
$f$ = 0.0, 1.0, and 1.3 [cf. Eq. (\ref{eq:T0surface})] while the 
spin-singlet pairing is fixed. The unperturbed pair 
transfer strength is also shown by a dotted line. In the lower panel, the 
experimental level scheme is inserted. This figure is based on the results of
Ref. \cite{Yoshida}.}
\label{fig:Yoshida-np}
\end{figure}

The proton-neutron $L=0$ pair transfer strength in the $N = Z$ nuclei 
$^{40}$Ca and $^{56}$Ni has been studied by K. Yoshida in Ref. 
\cite{Yoshida} by using the proton-neutron RPA described in 
Sec. \ref{Sect.03-1a} with a Skyrme EDF. The pair addition strength
of $^{40}$Ca $\rightarrow$ $^{42}$Sc associated with both 
the $J^{\pi}=1^+$ and $J^{\pi}=0^+$ states is shown in 
Fig. \ref{fig:Yoshida-np}. It was found that the collectivity of the lowest 
$J^{\pi}=1^+$ in the neighboring odd-odd nucleus $^{42}$Sc is stronger 
than that of the lowest $J^{\pi}=0^+$ state when the IS spin-triplet 
pairing is taken to be equal or stronger than the IV spin-singlet 
pairing.  

One sees that the excitation energy and the strength of
the $J^{\pi}=1^+$ states are strongly affected by the $T = 0$ pairing
interaction. In the case of $f$ = 0, without the $T = 0$
pairing interaction, the lowest $1^+$ state in $^{42}$Sc 
located at $\omega$= 7.5 MeV is a single-particle excitation 
$\pi f_{7/2} \nu f_{7/2}$. As the pairing interaction is switched on, 
and the strength is increased, the 1$^+$ state is shifted 
downwards in energy with the enhancement of the transition strength.
With increasing paring strength to $f$ = 1.0 or 1.3
the lowest 1$^+$ state is constructed by many particle-particle 
excitations involving $f_{5/2}$ and $p_{3/2}$ orbitals located above the
Fermi levels as well as the $\pi f_{7/2} \nu f_{7/2}$ excitation. It
is particularly worth noting that the hole-hole excitations from the 
$sd-$shell have an appreciable contribution to generate
this $T = 0$ proton-neutron pair-addition vibrational mode, indicating
$^{40}$Ca core-breaking. The strong collectivity is associated 
by a coherent phase of these configurations of pp and hh excitations.


\section{Summary and Future perspectives}
\label{Sect.04}
The superfluidity in nuclei was firstly pointed out in a milestone paper by A. Bohr, 
B.R. Mottelson and D. Pines in 1958. This historical paper has raised a strong impact 
on the study of nuclear superfluidity both from 
experimental and theoretical viewpoints. These studies have established solid evidences 
of superfluidity in nuclei such as
the odd-even staggering in the mass systematics, the large energy gaps in the spectra of 
even-even nuclei, the quenching of the moment of inertia associated with rotational 
spectra of rare-earth deformed nuclei, and the enhancement 
of fission of actinide nuclei. Theoretically, the HF plus BCS and HFB theories have been developed 
and successfully applied to calculate the effect of pairing interactions on these phenomena. 

It has also been pointed out that 
the pairing correlations will be much stronger in the lower density regime than at normal density.  
Nuclear matter calculations show the onset of the BCS-BEC crossover phenomenon in the 
low density regime, i.e., 
the spatial correlation of the Cooper pair is changing from a BCS-type behavior with a large 
coherence length at normal density to a BEC-type behavior with a compact coherence length.  
It was also pointed out that the weakly bound nuclei with halo or skin nature may 
show the strong di-nucleon correlations which give rise to a similar behaviour as in 
the BCS-BEC crossover 
when one studies the nucleon pairs as
a function 
of the distance between the core and the center of the two nucleons.  
 
So far, most of the studies of the pairing correlations have been concentrated on the 
isovector $T=1$ spin-singlet pairing interaction. On the other hand, 
the isoscalar $T=0$ spin-triplet 
pairing correlations could be much stronger than the $T=1$ pairing ones. The importance of 
$T=0$ spin-triplet pairing was already pointed out in the 1970s. There have been many 
discussions on possible signatures of the spin-triplet pairing correlations, such as 
the Wigner energy term in the mass formula, the enhancement of neutron-proton 
pair transfer cross sections, the inversion of $J=0^+$ and $J=1^+$ states, and the large 
enhancement of GT strength in the low-energy region of $N=Z$ and $N=Z+2$ nuclei. However, 
so far no convincing evidence has been found about the spin-triplet superfluidity in nuclei. 
This may be due to the large spin-orbit splittings in the nuclear mean field and the large neutron 
excess $N>Z$ in stable nuclei with $A>40$, which prevent 
the formation of spin-triplet Cooper pairs in nuclei.  

In our paper, 
we have pointed out that the inversions of $J=0^+$ and $J=1^+$ states in the odd-odd 
$N=Z$ nuclei can be considered as a manifestation of strong
spin-triplet pairing correlations. These nuclei also display strong 
magnetic dipole transitions 
that can be explained by the introduction of $T=0$ pairing. Another kind of spin-dependent
transition are the Gamow-Teller excitations from even-even $N=Z+2$ to the neighbouring
odd-odd ones. We have shown that the strong
concentration of GT strength at low energy, close to the ground state of daughter nuclei, 
can be understood by QRPA calculations only with a strong $T=0$ spin-triplet pairing interaction. 
In all these cases, the strength of the $T=0$ pairing is larger, or slightly larger, than 
the strength of the $T=1$ pairing, in agreement with similar conclusions extracted from
the analysis of the shell model matrix elements.

A direct evidence of the strong neutron-proton pairing correlations could appear in the enhancement 
of the pair transfer cross sections. Several theoretical studies show clear signs of the strong 
pairing correlations in the two-nucleon pairing strength. However, the reaction 
process of two-nucleon transfer reactions 
is very involved. While some calculations have been performed in the case of two-neutron
transfer, and are consistent with the assumptions on $T=1$ pairing coming from other pieces
of evidence, the study of proton-neutron transfer reactions is still in its infancy and more 
effort should be put on this topic.

The spin-triplet superfluidity is a fascinating subject for theoretical and experimental study.
In the future, it would be
highly interesting to build a realistic model able to predict the two-particle transfer cross 
sections theoretically, and point to possible spin-triplet 
pairing modes in the pair transfer reactions such as ($^3$He, p) or (p, $^3$He).
  
A large amplitude collective motion such as fission is also very sensitive to whether the nucleus is 
a viscous fluid or a superfluid. Qualitatively, it is definitely true that superfluidity 
enhances the fission probability significantly. The fission induced by low-energy excitations 
which occur in neutron capture reactions
are now feasible to be studied by microscopic 
theories like
time-dependent HFB. These theoretical studies can provide observables such as the 
internal energies of fission fragments which can be compared with experimental data.
We expect that this may also be instrumental for our understanding of pairing in nuclei.


\begin{ack}
We would like to thank Y. Fujita, K. Hagino, T. Sasano, Y. Tanimura, T. Uesaka, A. Vitturi
for fruitful discussions and comments. 
\end{ack}

\appendix
\section{Two-body matrix elements of $\delta -$type and separable pairing interactions}
\subsection{$\delta -$type pairing interaction}
We adopt the helicity representation to derive simple formulas involving only 
Clebsh-Gordan coeeficients \cite{BM2,Bertsch}. In the helicity representation, 
the single-particle wave function reads
\be
\psi_{nljm}=R_{nlj}(r)(\frac{2j+1}{16\pi^2})^{1/2}\sum_{h=\pm1/2}\alpha(ljh)D^{j}_{mh}(\hat{r})\chi_h, 
\ee
where $R_{nlj}(r)$, $D^{j}_{mh}$ and $\chi_h$ are the radial wave function and $D-$function 
in the helicity function, respectively, and 
\be
\alpha(ljh)=(-)^{(h+1/2)(j-l-1/2)}.
\ee
The two particle wave function can be written for the contact interaction 
$\delta(\vec r_1-\vec r_2)$ with $\hat{r_1}$=$\hat{r_2}$ as
\bea
&&\Psi(\alpha \beta)_{JM}=\frac{1}{\sqrt{1+\delta_{\alpha,\beta}}} \sum_{h_{\alpha},h_{\beta}}<j_{\alpha}h_{\alpha}j_{\beta}h_{\beta}|JH>D^{J}_{MH}(\hat{r})
 \nonumber
 \\
 && \times R_{\alpha}(r)R_{\beta}(r)  
 \frac{\hat{j}_{\alpha}\hat{j}_{\beta} } {16\pi^2}\alpha(l_{\alpha}j_{\alpha}h_{\alpha})\alpha(l_{\beta}j_{\beta}h_{\beta})\chi_{h_{\alpha}}
\chi_{h_{\beta}}, \nonumber
 \\
\label{h-2p}
\eea
where $\alpha\equiv(n_{\alpha}l_{\alpha}j_{\alpha})$ and $\hat{j}=(2j+1)^{1/2}$. To derive 
Eq. (\ref{h-2p}) we use the addition and the orthogonality relations of the $D-$ functions 
(see Ref. \cite{BM2} for details).
For the total spin $S=0$ state, the helicity $H$ is restricted to be $H=0$ only.
By using the helicity representaiton of the two-particle state, 
the two-body matrix element for the contact pairing for the isovector spin-singlet channel
$V^{(T=1,S=0)}(\br)=G^{(T=1)}f(\rho)\delta(\br)$ is calculated to be 
\begin{eqnarray*}
&&<\Psi(\alpha' \beta)'_{JM}|V^{(T=1,S=0)}|\Psi(\alpha \beta)_{JM}>= \nonumber \\
&&\frac{1}{\sqrt{1+\delta_{\alpha,\beta}}}\frac{1}{\sqrt{1+\delta_{\alpha',\beta'}}}
<j_{\alpha}1/2j_{\beta}-1/2|J0>\nonumber \\
& &<j_{\alpha}'1/2j_{\beta}'-1/2|J0> \frac{\hat{j}_{\alpha}\hat{j}_{\beta}\hat{j}_{\alpha}'\hat{j}_{\beta}'  } {16\pi (2J+1)} \nonumber\\
& &\times (-)^{(j_{\alpha}-l_{\alpha}-1/2+j_{\alpha}'-l_{\alpha}'-1/2)}  [1+(-)^{(l_{\alpha}+l_{\beta}+l_{\alpha}'+l_{\beta}')}] \nonumber \\
& &\times I^{(T=1)}, 
\end{eqnarray*}
where $I$ is the radial integral
\be
I^{(T=1)}=\int dr r^2 R_{\alpha}R_{\alpha'} R_{\beta}R_{\beta'}G^{(T=1)}f(\rho).
\label{radial-int}
\ee
Here the $T=1$ pairing strength is $G^{(T=1)}$ and the density dependent form factor is written 
as $f(\rho)$.  
For the anti-symmetrized state $\tilde{\Psi}$, the two body matrix elements will be
\bea
&&<\tilde{\Psi}(\alpha' \beta')_{JM}^T|V^{(T=1,S=0)}|\tilde{\Psi}(\alpha \beta)_{JM} ^T>= \nonumber \\
&&<\Psi(\alpha' \beta')_{JM}^T|V^{(T=1,S=0)}\{|\Psi(\alpha \beta)_{JM}^T \nonumber \\
&&-(-)^{j_{\alpha}+j_{\beta}-J-T-1}|\Psi(\beta \alpha )_{JM}^T>\}=  \nonumber \\
&&\frac{1}{\sqrt{1+\delta_{\alpha,\beta}}}\frac{1}{\sqrt{1+\delta_{\alpha',\beta'}}}
\times<j_{\alpha}1/2j_{\beta}-1/2|J0>\nonumber \\
& &<j_{\alpha}'1/2j_{\beta}'-1/2|J0> \frac{\hat{j}_{\alpha}\hat{j}_{\beta} } {8\pi} \frac{\hat{j}_{\alpha}'\hat{j}_{\beta}' } {(2J+1)} \nonumber\\
& &\times (-)^{(j_{\alpha}-l_{\alpha}-1/2+j_{\alpha}'-l_{\alpha}'-1/2)}  [1+(-)^{(l_{\alpha}+l_{\beta}+J)}] \nonumber \\
& &\times I^{(T=1)}.
\label{eq:T=1ame}
\eea

For the spin-triplet case, we have to calculate not only the $H=0$ state, but also $H=\pm 1$ states.
The $H=$0 state give the same 2-body matrix element as the spin-singlet case (\ref{eq:T=1ame})  except for the $J$ selection 
term which will be $[1-(-)^{(l_{\alpha}+l_{\beta}+J)}]$ in the $S=1$ case. We will present 
how to calculated the matrix element for the $H=-1$ state in the following. 
The matrix element for $H=1$ is essentially the same as that of $H=-1$ state. 
The two-particle state for the total helicity $H=-1$ reads 
\bea
\Psi(\alpha \beta)_{JM}&=&\frac{1}{\sqrt{1+\delta_{\alpha,\beta}}} <j_{\alpha}-1/2j_{\beta}-1/2|J-1>
 \nonumber
 \\
  &\times&    D^{J}_{M-1}(\hat{r}) R_{\alpha}(r)R_{\beta}(r)  
  \frac{\hat{j}_{\alpha}\hat{j}_{\beta} } {16\pi^2})\chi_{-1/2}
\chi_{-1/2},   \nonumber
 \\
\label{h-2p-1}
\eea
where $\alpha(ljh)$ in Eq. (\ref{h-2p}) becomes always $+1$ for $h=-1/2$ state.
The two-body matrix element of the $S=1, H=-1$ state for the isoscalar spin-triplet interaction 
$V^{(T=0,S=1)}=G^{(T=0)}f(\rho)\delta(\br)$ will be
  \begin{eqnarray*}
&&<\Psi(\alpha' \beta)'_{JH=-1}|V^{(T=0,S=1)}|\Psi(\alpha \beta)_{JH=-1}>= \nonumber \\
&&\frac{1}{\sqrt{1+\delta_{\alpha,\beta}}}\frac{1}{\sqrt{1+\delta_{\alpha',\beta'}}}
<j_{\alpha}-1/2j_{\beta}-1/2|J-1>\nonumber \\
& &<j_{\alpha}'-1/2j_{\beta}'-1/2|J-1> \frac{\hat{j}_{\alpha}\hat{j}_{\beta}\hat{j}_{\alpha}'\hat{j}_{\beta}'  } {16\pi (2J+1)} \nonumber\\
& &\times (-)^{j_{\alpha}+j_{\alpha}'+j_{\beta}+j_{\beta}'}  \times I^{(T=0)} ,   \\
\label{2b-S=1}
\end{eqnarray*}
where the radial integral $I^{(T=0)}$ has the pairing strength $G^{(T=0)}$ instead of $G^{(T=1)}$ 
in Eq. (\ref{radial-int}).  
This formula (\ref{2b-S=1}) can be rewritten further by using a recursion formula
\bea
&&<j_{\alpha}-1/2j_{\beta}-1/2|J-1>=<j_{\alpha}1/2j_{\beta}-1/2|J0> \nonumber \\
&&\times \frac{(-)^{j_{\alpha}+j_{\beta}-J}
(j_{\alpha}+1/2)+(j_{\beta}+1/2)}{[J(J+1)]^{1/2}}.
\eea
Then, the two-body matrix element becomes
\begin{eqnarray*}
&&<\Psi(\alpha' \beta)'_{JH=-1}|V^{(T=0,S=1)}|\Psi(\alpha \beta)_{JH=-1}>= \nonumber \\
&&\frac{1}{\sqrt{1+\delta_{\alpha,\beta}}}\frac{1}{\sqrt{1+\delta_{\alpha',\beta'}}}
<j_{\alpha}1/2j_{\beta}-1/2|J0> \nonumber \\
& &\times<j_{\alpha}'1/2j_{\beta}'-1/2|J0> \nonumber \\
&&\times \frac{(-)^{j_{\alpha}+j_{\beta}-J}
(j_{\alpha}+1/2)+(j_{\beta}+1/2)}{[J(J+1)]^{1/2}}\nonumber \\
&&\times \frac{(-)^{j_{\alpha}'+j_{\beta}'-J}
(j_{\alpha}'+1/2)+(j_{\beta}'+1/2)}{[J(J+1)]^{1/2}} \frac{\hat{j}_{\alpha}\hat{j}_{\beta}\hat{j}_{\alpha}'\hat{j}_{\beta}'  } {16\pi (2J+1)} \\
& &\times (-)^{j_{\alpha}+j_{\alpha}'+j_{\beta}+j_{\beta}'}  \times I^{(T=0)}   .
\label{2b-S=1a}
\end{eqnarray*}
We have to sum up three terms $H=0, H=\pm1$ for the matrix element of the $S=1$ state. 
We remind that $H=+1$ state gives 
the same matrix element as that of $H=-1$ state. Eventually, for the anti-symmetrized $S=1$ state, the
two-body matrix element is given by
\bea
&&<\tilde{\Psi}(\alpha' \beta')_{JM}^{(T=0)}|V^{(T=0,S=1)}|\tilde{\Psi}(\alpha \beta)_{JM}^{(T=0)}>=\nonumber \\  
&&\frac{1}{\sqrt{1+\delta_{\alpha,\beta}}}\frac{1}{\sqrt{1+\delta_{\alpha',\beta'}}}
\times<j_{\alpha}1/2j_{\beta}-1/2|J0>\nonumber \\
&&<j_{\alpha}'1/2j_{\beta}'-1/2|J0> \frac{\hat{j}_{\alpha}\hat{j}_{\beta} } {8\pi} \frac{\hat{j}_{\alpha}'\hat{j}_{\beta}' } {(2J+1)} \nonumber\\
&&\times (-)^{j_{\alpha}-l_{\alpha}-1/2+j_{\alpha}'-l_{\alpha}'-1/2}   [(1+(-)^{l_{\alpha}+l_{\beta}+J+1})
\nonumber \\
&&+ 2(-)^{j_{\beta}+j_{\beta}'+l_{\alpha}+l_{\alpha}'+1} 
\frac{(-)^{j_{\alpha}+j_{\beta}-J}
(j_{\alpha}+1/2)+(j_{\beta}+1/2)}{[J(J+1)]^{1/2}}\nonumber \\
&&\times \frac{(-)^{j_{\alpha}'+j_{\beta}'-J}
(j_{\alpha}'+1/2)+(j_{\beta}'+1/2)}{[J(J+1)]^{1/2}}]\nonumber \\
&&\times I^{(T=0)} . 
\eea

\begin{table}  [b]
 \caption{ \label{tab:9j}
The transformation coefficient $R$ between the 
$jj$ coupling and the $LS$ coupling for the pair wave functions,
 $R=\langle[(l\frac{1}{2})^{j}(l\frac{1}{2})^{j'}]^{J=1}|[(ll)^{L=0}(\frac{1}{2}\frac{1}{2})^{S=1}]^{J=1}\rangle$.
$\Omega$ is defined as $\Omega\equiv 3(2l+1)^2$ 
.}
\begin{tabular}{cc|c|c|c}  \hline
 $j$ &$j'$ & $R$ & $l=1$  & $l=3$  \\\hline
$l+1/2$ &$l+1/2$  &  $\sqrt{\frac{(2l+2)(2l+3)}{2\Omega}}$   & $\frac{1}{3}\sqrt{\frac{10}{3}}$ & $\frac{2\sqrt{3}}{7}$
    \\
$l+1/2$ &$l-1/2$  &  $-\sqrt{\frac{4l(l+1)}{\Omega}}$   & $-\frac{2}{3}\sqrt{\frac{2}{3}}$  & $-\frac{4}{7}$ \\
 $l-1/2$ & $l-1/2$  & $-\sqrt{\frac{2l(2l-1)}{2\Omega}}$   & $-\frac{1}{3}\sqrt{\frac{1}{3}}$ & $-\frac{\sqrt{5}}{7}$ \\
 $l-1/2$  & $l+1/2$ & $\sqrt{\frac{4l(l+1)}{\Omega}}$   & $\frac{2}{3}\sqrt{\frac{2}{3}}$  & $\frac{4}{7}$ \\ \hline
\end{tabular}
\end{table}

\subsection{Separable pairing interaction}
Let us discuss next the separable pairing interaction, i.e.,   
the spin-singlet $T=1$ pairing interaction (\ref{eq:T=1}) and 
the spin-triplet $T=0$ pairing interaction (\ref{eq:T=0}). 
The two-body matrix element for the $T=1$ pairing is evaluated to be
\bea
\langle (j_ij_i)T=1,J=0|V^{(T=1)}|(j_jj_j)T=1,J=0\rangle \nonumber \\
=-\sqrt{(j_i+1/2)(j_j+1/2)}\,G^{(T=1)}I_{ij}^2,
\label{eq:T=1me}
\eea
where $I_{ij}$ is the overlap integral given by 
\bea \label{overlap}
I_{ij}=\int\psi_i(\br)^*\psi_j(\br)d\br.  
\eea
For the $T=0$ pairing, the two-body matrix element 
involves the coefficient for the transformation from 
the $jj$ coupling scheme to 
the $LS$ coupling scheme, and is given by 
\begin{eqnarray*}
&&\langle (j_1j_2)T=0,J=1|V^{(T=0)}|(j_1'j_2')T=0,J=1\rangle=\nonumber\\
&-&\left\langle\left[\left(l_1\frac{1}{2}\right)^{j_1}\left(l_2\frac{1}{2}\right)
^{j_2}\right]^{J=1}\left|\left[
\left(l_1l_2\right)^{L=0}\left(\frac{1}{2}\frac{1}{2}
\right)^{S=1}\right]^{J=1}\right.\right\rangle \nonumber \\
&\times&\left\langle\left[\left(l_1'\frac{1}{2}\right)^{j_1'}\left(l_2'
\frac{1}{2}\right)
^{j_2'}\right]^{J=1}\left|\left[
\left(l_1'l_2'\right)^{L=0}\left(\frac{1}{2}\frac{1}{2}
\right)^{S=1}\right]^{J=1}\right.\right\rangle \nonumber \\
&\times&\frac{\sqrt{2l_1+1}\sqrt{2l_{1}'+1}}{\sqrt{1+\delta_{j_1,j_2}}\sqrt{1+\delta_{j_1',j_2'}}}\,
 fG^{(T=1)}(I_{j_1j_1'}I_{j_2j_2'}+I_{j_1j_2'}I_{j_1j_2'}), 
\label{eq:T=0me}
\end{eqnarray*}
where 
$\langle[(l_1\frac{1}{2})^{j_1}(l_2\frac{1}{2})^{j_2}]^{J=1}|[(l_1l_2)^{L=0}(\frac{1}{2}\frac{1}{2})^{S=1}]^{J=1}\rangle$ 
is the transformation coefficient, and 
the overlap integral $I_{ij}$ involves 
both the proton and neuron wave functions.
The transformation coefficient can be evaluated 
with the 9j symbol [cf. Eq. (\ref{eq:9jtrasf})] and the explicit form is summarized 
in Table \ref{tab:9j}. 
The square of the transformation coefficient is 1/6 and 1/3 
for $j_1=j_2$ and $j_1=j_2\pm1$ configurations, respectively, in the limit
of large angular momentum $l\rightarrow\infty$. 
These values suggest a large quenching of the spin-triplet 
pairing correlations, and that 
spin-orbit partners contribute largely to 
the spin-triplet pairing matrix elements.  
On the other hand, in the small $l$ limit, $l\rightarrow 0$, 
the  coefficient is unity for $j=j'=l+1/2$, and 
the coefficients are zero for the other three 
configurations.  
This suggests that 
the spin-triplet pairing is as large as 
the spin-singlet pairing for the pair configuration in the $s_{1/2}$ orbit, 
and that it is still 
substantially large for the configuration in the $p_{3/2}$ orbit. 
%


\end{document}